\documentclass[aps,prd,amsmath,amssymb,reprint,superscriptaddress]{revtex4-2}

\usepackage{graphicx} 
\usepackage{dcolumn}  
\usepackage{booktabs}
\usepackage{bm}       
\usepackage[usenames,dvipsnames]{xcolor}
\usepackage{hyperref} 
\usepackage[normalem]{ulem}
\hypersetup{
    colorlinks = true,
    linkcolor = Blue,
    citecolor = Blue,
    urlcolor = Blue
}

\newcommand{\orcid}[1]{\href{https://orcid.org/#1}{\includegraphics[width=0.7em]{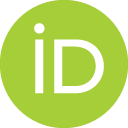}}}
\newcommand{\araa}{Annu. Rev. Astron. Astrophys.}
\newcommand{\apjl}{Astrophys. J. Lett.}
\newcommand{\apjs}{Astrophys. J. Suppl.}
\newcommand{\aap}{Astron. Astrophys.}
\newcommand{\jcap}{J. Cosmol. Astropart. Phys.}
\newcommand{\mnras}{Mon. Not. R. Astron. Soc.}

\newcommand{\physrep}{Phys. Rep.}

\newcommand{\datavh}{\bm{\hat{\mathcal{D}}}}
\newcommand{\datav}{\bm{\mathcal{D}}}
\newcommand{\covc}{\bm{\mathcal{C}}}
\newcommand{\deltaK}{\delta^{\rm K}}
\newcommand{\deltaD}{\delta^{\rm D}}
\newcommand{\bl}{\boldsymbol{l}}

\begin{document}

\title{Joint analyses of 2D CMB lensing and 3D galaxy clustering in the spherical Fourier-Bessel basis}


\author{Yucheng Zhang \orcid{0000-0002-9300-2632}}
\email[]{yucheng.zhang@nyu.edu}
\affiliation{Center for Cosmology and Particle Physics, Department of Physics, New York University, 726 Broadway, New York, NY 10003, USA}

\author{Anthony R. Pullen}
\affiliation{Center for Cosmology and Particle Physics, Department of Physics, New York University, 726 Broadway, New York, NY 10003, USA}
\affiliation{Center for Computational Astrophysics, Flatiron Institute, 162 Fifth Ave, New York, NY 10010, USA}

\author{Abhishek S. Maniyar}
\affiliation{Center for Cosmology and Particle Physics, Department of Physics, New York University, 726 Broadway, New York, NY 10003, USA}

\date{\today}

\begin{abstract}
Cross-correlating cosmic microwave background (CMB) lensing and galaxy clustering has been shown to greatly improve the constraints on the local primordial non-Gaussianity (PNG) parameter $f_{\rm NL}$ by reducing sample variance and also parameter degeneracies.
To model the full use of the 3D information of galaxy clustering, we forecast $f_{\rm NL}$ measurements using the decomposition in the spherical Fourier-Bessel (SFB) basis, which can be naturally cross-correlated with 2D CMB lensing in spherical harmonics.
In the meantime, such a decomposition would also enable us to constrain the growth rate of structure, a probe of gravity, through the redshift-space distortion (RSD).
As a comparison, we also consider the tomographic spherical harmonic (TSH) analysis of galaxy samples with different bin sizes.
Assuming galaxy samples that mimic a few future surveys, we perform Fisher forecasts using linear modes for $f_{\rm NL}$ and the growth rate exponent $\gamma$, marginalized over standard $\Lambda$ cold dark matter ($\Lambda$CDM) cosmological parameters and two nuisance parameters that account for clustering bias and magnification bias.
Compared to TSH analysis using only one bin, SFB analysis could improve $\sigma(f_{\rm NL})$ by factors 3 to 12 thanks to large radial modes.
With future wide-field and high-redshift photometric surveys like the LSST, the constraint $\sigma(f_{\rm NL}) < 1$ could be achieved using linear angular multipoles up to $\ell_{\rm min}\simeq 20$.
Compared to using galaxy auto-power spectra only, joint analyses with CMB lensing could improve $\sigma(\gamma)$ by factors 2 to 5 by reducing degeneracies with other parameters, especially the clustering bias.
For future spectroscopic surveys like the DESI or \textit{Euclid}, using linear scales, $\gamma$ could be constrained to $3\,\%$ precision assuming the GR fiducial value.
\end{abstract}

\maketitle

\section{\label{sec:intro}Introduction}
In many large-scale cosmological surveys, the observables can be classified as tracers of the matter field, the main ingredient of which is the invisible and mysterious dark matter~\cite{PDG2020} that is known to be interacting with baryonic matter through gravity.
The 3D large-scale structure (LSS) of the matter field can be traced with photons emitted directly from baryonic matter, e.g. in galaxy redshift and line intensity mapping (LIM) surveys (see e.g.~\cite{eBOSS2021,Schaan2021}).
On the other hand, a 2D map of the line-of-sight (LOS) integral of the matter field can also be reconstructed up to a distant light source through the weak gravitational lensing effect~\cite{Hoekstra2008}.
The light source can be luminous matter at different redshifts~\cite{Mandelbaum2018} or cosmic microwave background (CMB) traveling from the epoch of recombination~\cite{Lewis2006}.

The lensing convergence signal reconstructed from CMB temperature and polarization maps is the LOS integral of the matter field up to redshift $z\sim 1100$, and therefore should correlate with any galaxy clustering observations.
These cross-correlations have been detected in several previous works using different CMB lensing and galaxy clustering datasets, see e.g.~\cite{Smith2007,Hirata2008} for the first two detections.
With the cross-correlation, CMB lensing and galaxy clustering can also be further combined to construct other statistics like $E_G$~\cite{Zhang2007,Pullen2015,Pullen2016,Singh2019,Zhang2020} to probe gravity.

Joint analysis of CMB lensing and galaxy clustering has been shown to be powerful in improving the constraint on the local primordial non-Gaussianity (PNG) parameter $f_{\rm NL}$~\cite{Seljak2009,Schmittfull2018,Ballardini2019,Chen2021,Climent2021}.
CMB lensing is an unbiased tracer of the matter field, while galaxy clustering has a bias that could be scale-dependent due to PNG~\cite{Dalal2008,Slosar2008}.
This difference in bias of the two tracers makes the joint analysis useful in reducing the sample variance and mitigating the degeneracies between $f_{\rm NL}$ and other cosmological parameters.

In previous joint analyses, the galaxies in a redshift bin are usually projected in the radial direction to make an angular map to be cross-correlated with the CMB lensing map, typically in spherical harmonic (SH) space.
However, the radial information of the 3D galaxy field could be lost in the projection.
Even if we split the redshift coverage of a galaxy sample into many bins and perform the tomographic spherical harmonic (TSH) analysis with the covariances between redshift bins fully included, it is still uncertain how well the radial information could be recovered for different scales that are mixed, see e.g.~\cite{Taylor2021} for a recent discussion on this.
The standard 3D analysis of galaxy clustering is usually based on the Cartesian Fourier transform.
However, this makes it difficult to do the cross-correlation with the SH coefficients of 2D angular maps given the different bases.
Also for analysis in Cartesian coordinates, large scales are quite challenging given the spherical geometric boundaries of the survey and also LOS effects like redshift-space distortions (RSD)~\cite{Castorina2020}.

The positions of the galaxies are measured in spherical coordinates, for which the spherical Fourier-Bessel (SFB) decomposition would be a natural choice for power spectrum analyses.
SFB analysis decomposes a 3D field in the spherical eigenfunctions of the Laplacian, which are spherical harmonics and spherical Bessel functions.
There have been a number of studies about SFB analysis of galaxy clustering, which can be traced back to~\cite{Fisher1995,Heavens1995}.
Here we list some of the recent discussions with further references cited therein.
\cite{Samushia2019} suggested the proper radial basis function to be used in spherical shells, which is a more optimal choice for surveys that do not start from redshift zero.
\cite{Leistedt2012,Gebhardt2021} developed the SFB power spectrum estimators.
\cite{Lanusse2015} compared SFB and tomographic analyses in parameter constraints, and found that SFB analysis is more robust to systematics in galaxy clustering bias.
\cite{Passaglia2017} discussed cross-correlations of 2D photometric and 3D spectroscopic galaxy surveys.
\cite{Wang2020} proposed a hybrid-basis inference by combining SFB and Cartesian Fourier analyses on different scales.
Besides galaxy clustering, SFB formalism has also been discussed for LIM in power spectrum analysis~\cite{Liu2016} and full sky lensing reconstruction~\cite{Chakraborty2019}.

In this work, we consider the joint analyses of 2D CMB lensing and 3D galaxy clustering, which are decomposed in SH and SFB bases, respectively.
The same angular basis function makes it straightforward to cross-correlate 2D and 3D fields using their SH and SFB coefficients.
In this SFB formalism of galaxy clustering, we discuss the expressions for power spectra, including modifications due to PNG, RSD and also magnification bias.
Then we perform Fisher forecasts for the constraints on $f_{\rm NL}$ and the growth rate exponent $\gamma$, with a set of standard $\Lambda$ cold dark matter ($\Lambda$CDM) cosmological parameters and two nuisance parameters accounting for galaxy clustering bias and magnification bias being marginalized.
We assume a few galaxy sample setups that mimic the designed specifications of some future spectroscopic and photometric surveys, including the Dark Energy Spectroscopic Instrument (DESI)~\cite{web-desi}, the \textit{Euclid} satellite mission~\cite{web-euclid}, the Legacy Survey of Space and Time (LSST)~\cite{web-lsst} of the Vera C. Rubin Observatory, and the Spectro-Photometer for the History of the Universe, Epoch of Reionization, and Ices Explorer (SPHEREx)~\cite{web-spherex}.
We consider only linear modes that are quantified with SFB and TSH power spectra directly, which we show to be better defined than converting a 3D wavenumber to an angular mode in 2D harmonic space, as is typically done.
For $f_{\rm NL}$, it would be interesting to check the improvement with large radial scales, which should not only contribute more information but also help in reducing sample variance.
Thanks to the SFB transform in fully including radial information on all scales, we are able to constrain $\gamma$ simultaneously.
For these two parameters that appear in galaxy clustering only, we investigate how CMB lensing could contribute to the constraints by mitigating the degeneracies with other cosmological or nuisance parameters.
As a comparison to SFB, we also consider the TSH analysis of the galaxy samples, and study how the information from linear modes depend on different bin sizes.

The paper is organized as follows.
First we briefly review the CMB lensing and galaxy number density fields in Section~\ref{sec:tracer}, where modifications to the galaxy field due to PNG, RSD and magnification bias are also discussed.
The angular SH and 3D SFB decomposition and power spectra of these fields are described in Section~\ref{sec:power}, including the well-known noises in auto-power spectra.
In Section~\ref{sec:surveys}, details of fiducial CMB lensing and galaxy redshift surveys are introduced.
For these surveys, we perform Fisher forecasts on parameter constraints, with the setup described in Section~\ref{sec:fisher}.
We present and discuss the results in Section~\ref{sec:results}, and conclude in Section~\ref{sec:conclusions}.
In this work, we assume a flat $\Lambda$CDM cosmology with \textit{Planck} 2018 CMB TT,TE,EE+lowE best-fit parameters~\cite{Planck2018-cosmo} as fiducial values.

\section{\label{sec:tracer}Tracers of the matter field}
In this section, we briefly review the observables in CMB lensing and galaxy redshift surveys, and their connection to the matter field.

Since the photons take a finite time to travel to us, we are actually observing the past light cone instead of the 3D matter field at $z=0$.
For both CMB lensing and galaxy surveys, the matter field traced at radial comoving distance $r(z) = \int_0^z dz'\,c/H(z')$ is the status of the field at redshift $z$, which uniquely corresponds to the time that the light was emitted.
In linear perturbation theory, the redshift evolution of the matter field can be described with
\begin{equation} \label{eq:delta_m_evo}
    \delta_m(\bm{r},z) = D(z) \delta_{m,0}(\bm{r}) \,,
\end{equation}
where $\delta_m(\bm{r},z)$ is the 3D matter field at redshift $z(r)$, $\bm{r}\equiv (r,\hat{r})$ with $\hat{r}\equiv (\theta, \phi)$ denoting the angular coordinates, $D(z)$ is the linear growth factor normalized to $D(z=0)=1$, and $\delta_{m,0}(\bm{r})$ denotes the 3D matter field at redshift $z=0$.

\subsection{\label{subsec:cmb_lensing}CMB lensing map}
The CMB lensing signal reconstructed from CMB temperature and polarization maps traces the integral of the matter field along the line-of-sight direction
\begin{equation} \label{eq:cmb_lensing_kappa}
    \kappa(\hat{r}) = \int_0^{r_{\rm CMB}} dr\, W_{\kappa}(r, r_{\rm CMB}) \delta_m(\bm{r},z) \,,
\end{equation}
where $r_{\rm CMB}$ is $r$ at redshift $z_{\rm CMB}\simeq 1100$, and the lensing kernel
\begin{equation} \label{eq:lensing_kernel}
    W_{\kappa}(r, r_\star) = \frac{3\Omega_{m,0}H_0^2}{2c^2}(1+z)r\left(1-\frac{r}{r_\star}\right) \,,
\end{equation}
with the light source being CMB and located at $r_\star=r_{\rm CMB}$ in this case.

\subsection{\label{subsec:galaxy}Galaxy clustering catalog}

\subsubsection{Number density}
LSS galaxy or quasar surveys construct catalogs that include the angular positions and redshifts of a large number of point sources that are selected for clustering analyses.
Assuming uniform angular target selection, the number density field can be written as
\begin{equation} \label{eq:n_3D}
    n(r,\hat{r}) = \bar{n}_V\phi_g(r)\left[1+\delta_g(r,\hat{r})\right] \,,
\end{equation}
where the average volume number density $\bar{n}_V=N/V$ is given by the ratio of the total number of targets and the comoving volume of the survey, $\phi_g(r)$ is the radial selection function of the survey, and $\delta_g(r,\hat{r})$ is related to the matter field through
\begin{equation} \label{eq:delta_g_m}
    \delta_g(r,\hat{r}) = b_g(z) \delta_m(\bm{r},z) \,,
\end{equation}
where $b_g(z)$ is the galaxy clustering bias, which is usually redshift-dependent and can also be scale-dependent (e.g. due to primordial non-Gaussianity, discussed below).
Given the number density field constructed from the catalogs, we can define an overdensity field
\begin{equation} \label{eq:delta_data}
    \delta(r,\hat{r}) \equiv \frac{n(r,\hat{r}) - \bar{n}_V}{\bar{n}_V} \,,
\end{equation}
and leave $\phi_g$ in the relation to $\delta_g$,
\begin{equation} \label{eq:delta_model}
    \delta(r,\hat{r}) = \phi_g(r)\delta_g(r,\hat{r}) + \phi_g(r) - 1 \,.
\end{equation}
To get $\phi_g(r)$ for the survey, we first define the normalized redshift distribution
\begin{equation}
    f_g(z) \equiv \frac{1}{N}\frac{dN}{dz} \,,
\end{equation}
which can be directly constructed with all the redshifts in the catalog, e.g. by making a histogram. The relation between $\phi_g(r)$ and $f_g(z)$ can be derived by considering the number of targets in a thin radial slice
\begin{equation}
    \int_{\Omega}d\Omega\, r^2dr\, n(r,\hat{r}) = N f_g(z) dz \,,
\end{equation}
which gives
\begin{equation} \label{eq:phig_fg}
    \bar{n}_V\phi_g(r) r^2dr = \bar{n}_\Omega f_g(z)dz \,,
\end{equation}
where $\bar{n}_\Omega = N/\Omega$ is the average angular number density, i.e. the number of targets per solid angle.
With this relation, $\phi_g(r)$ and $f_g(z)$ can be used interchangeably in describing the radial distribution of galaxies.

Similar as the 3D field $\delta(r,\hat{r})$, the projected 2D galaxy overdensity map $g(\hat{r})$ is usually constructed as
\begin{equation} \label{eq:g_data}
    g(\hat{r}) \equiv \frac{\int n(r,\hat{r}) r^2 dr - \bar{n}_\Omega}{\bar{n}_\Omega} \,.
\end{equation}
Combining Eq.~(\ref{eq:n_3D}) and~(\ref{eq:phig_fg}), $g$ is related to $\delta_g$ through
\begin{equation} \label{eq:g_model}
    g(\hat{r}) = \int dr\, \frac{H(z)}{c} f_g(z) \delta_g(r,\hat{r}) \,,
\end{equation}
where $dr=c\,dz/H(z)$ has been used.

To summarize, \textit{uniform} 3D galaxy overdensity fields can be simply connected to the matter perturbation field with Eq.~(\ref{eq:delta_g_m}).
However, due to the target selections in real surveys and depending on how the fields are constructed given the data, additional calibration functions like $\phi_g(r)$ or $f_g(z)$ may have to be applied.
For the 3D and 2D overdensity field constructed from observed catalogs using Eq.~(\ref{eq:delta_data}) and~(\ref{eq:g_data}), $\phi_g(r)$ and $f_g(z)$ are included in the corresponding theoretical modeling, Eq.~(\ref{eq:delta_model}) and~(\ref{eq:g_model}).
Of course, if we change how the fields were constructed from data, these analytic modelings would have to be modified accordingly.

\subsubsection{Redshift-space distortion}
The observed galaxy redshifts include contributions from not only the Hubble expansion but also the peculiar velocities of the targets due to gravity.
This causes a radial distortion (i.e. RSD) in the observed galaxy field compared with the true field.
In linear perturbation theory, the modification to $\delta_g(r,\hat{r})$ due to RSD can be described with~\cite{Kaiser1987,Hamilton1992,Hamilton1998}
\begin{equation}
    \Delta \delta_g(r,\hat{r})|_{\rm RSD} = f(z) \mathcal{R} \delta_m(\bm{r},z) \,,
\end{equation}
where $f(z)=d\ln D(z)/d\ln a$ is the linear growth rate defined as the logarithmic derivative of the growth factor with respect to the scale factor, and the RSD operator $\mathcal{R} \simeq \partial^2/\partial r^2\,\nabla^{-2}$, which results in a second-order derivative of the spherical Bessel function in the LOS integral, as we will see below.
In GR and some modified gravity models, the linear growth rate depends on the matter fraction through $f(z) = \Omega_m(z)^\gamma$~\cite{Linder2005}.
The exponent $\gamma\simeq 0.55$ for GR, and this value could vary for different gravity models,

\subsubsection{Primordial non-Gaussianity of local type}
\begin{figure}
    \centering
    \includegraphics[width=\columnwidth]{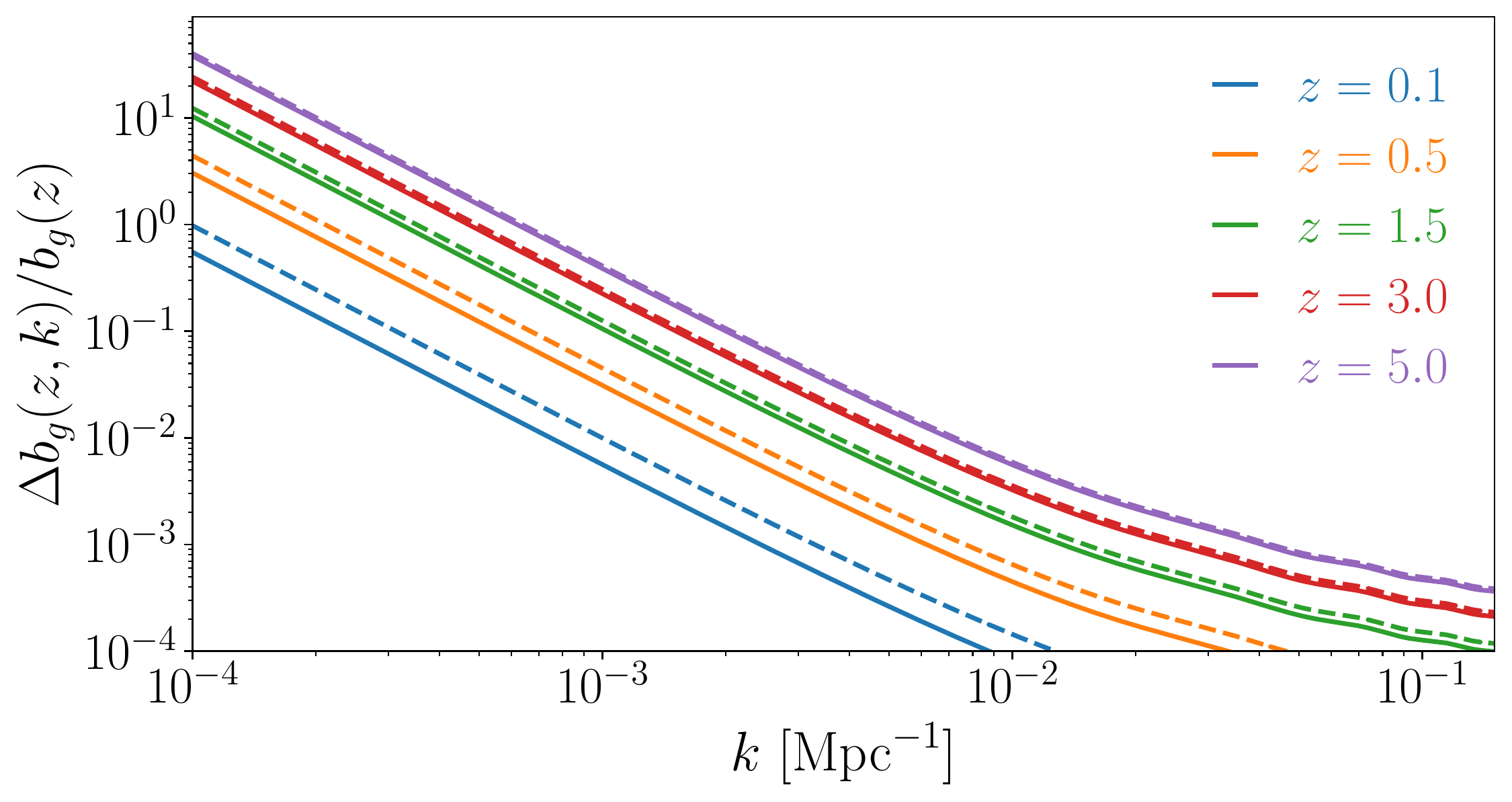}
    \caption{\label{fig:fnl_Delta_bg}Redshift and scale dependence of the clustering bias correction due to PNG of local type, shown as the fractional change to the fiducial bias $b_g(z)$ assuming $f_{\rm NL}=1$. Lines are plotted for $b_g(z)=1/D(z)$ (solid) or $b_g(z)=1+z$ (dashed), which represent two typical types of redshift dependence.}
\end{figure}

Measuring PNG is one of the promising methods to constrain models of inflation in the early universe, which sources the primordial density fluctuations and therefore the LSS of the matter field observed today.
PNG of local type is introduced to the primordial Gaussian potential $\psi(\bm{x})$ through $\Psi_{\rm NG}(\bm{x}) = \psi(\bm{x}) + f_{\rm NL}(\psi^2(\bm{x})-\langle\psi^2\rangle)$, with the non-Gaussian term proportional to the scale-independent $f_{\rm NL}$ parameter.
The standard single-field slow-roll inflation predicts a $f_{\rm NL}$ that is smaller than unity, while in other models like multifield inflation, $f_{\rm NL}$ could be significantly higher (see e.g.~\cite{Slosar2008} and references therein).

It was found that this local PNG leaves fingerprint on LSS tracers like galaxies by introducing a scale-dependent modification to the clustering bias~\cite{Dalal2008}
\begin{equation}
    b_g(z, k) = b_g(z) + f_{\rm NL} \Delta b_g(z, k) \,,
\end{equation}
where
\begin{equation}
    \Delta b_g(z, k) = 3 \left(b_g(z)-1\right) \frac{\Omega_{m,0}\delta_c}{k^2T(k)\tilde{D}(z) }\left(\frac{H_0}{c}\right)^2 \,,
\end{equation}
with $\tilde{D}(z)$ being the linear growth factor normalized to $(1+z)^{-1}$ for matter domination, i.e. $\tilde{D}(z) = \tilde{D}(0)D(z)$.
This bias correction is more significant on larger scales ($\propto k^{-2}$) and higher redshifts ($\propto D(z)^{-1}$), with a simple illustration in Fig.~\ref{fig:fnl_Delta_bg}.

\subsubsection{Magnification bias}
Just as the CMB photons are lensed by the matter field all the way from the last scattering surface to us, our observed galaxies are also lensed by the foreground matter field.
This weak gravitational lensing could change the flux of an individual target and also magnify the angular distribution of the targets.
The corresponding distortion to $\delta_g(r,\hat{r})$ is given by~\cite{Hui2007,Yang2018}
\begin{equation} \label{eq:mag_bias}
    \Delta\delta_g(r, \hat{r})|_{\rm lensing} = (5s-2) \kappa_g(r, \hat{r}) \,,
\end{equation}
where $s$ is the magnification bias parameter and the weak lensing convergence up to $r$ is given by
\begin{equation}
    \kappa_g(r, \hat{r}) = \int_0^r dr'\, W_\kappa(r', r) \delta_m(\bm{r}', z') \,,
\end{equation}
where the lensing kernel $W_\kappa$ is given in Eq.~(\ref{eq:lensing_kernel}), with the first and second parameter being the comoving distance to the lensing and light source respectively.
A subscript $g$ is added just to distinguish it from $\kappa$, which specifically refers to the CMB lensing in this paper.
For galaxy samples with a faint-end cutoff target selection, the magnification bias is given by~\cite{Hui2007}
\begin{equation} \label{eq:s_def}
    s = \left.\frac{d \log_{10}N(m<m_*)}{d m}\right|_{m=m_*} \,,
\end{equation}
where $m$ is the apparent magnitude, and $N(m<m_*)$ is the number of targets that appear to be brighter than the survey faint limit $m_*$.

\section{\label{sec:power}Power spectra}
In this section, we start with a brief review on the spherical Fourier analyses of 2D and 3D fields, which also defines the conventions of transforms in this work.
Applying these decompositions to the CMB lensing and galaxy overdensity fields, we derive the formalism for the auto- and cross-power spectra.

\subsection{\label{subsec:decomposition}Fourier decomposition in spherical coordinates}
A 2D field $a(\hat{r})$ defined on a sphere can be decomposed in spherical harmonic space as
\begin{equation} \label{eq:SH_trans}
    a(\hat{r}) = \sum_{\ell m} a_{\ell m} Y_{\ell m}(\hat{r}) \,,
\end{equation}
where $Y_{\ell m}(\hat{r})$ are the spherical harmonics that are orthonormal $\int d\Omega\,Y_{\ell m}(\hat{r})Y_{\ell' m'}(\hat{r})=\deltaK_{\ell\ell'}\deltaK_{mm'}$ by definition.
The coefficients are given by the inverse transform
\begin{equation} \label{eq:SH_coef}
    a_{\ell m} = \int d\Omega\, a(\hat{r}) Y_{\ell m}^*(\hat{r}) \,,
\end{equation}
with $d\Omega = \sin\theta d\theta d\phi$ being the differential solid angle.

A 3D field $f(r,\hat{r})$ expressed in spherical coordinates can be similarly decomposed in the SFB basis, which is a natural extension to the angular case above with the radial coordinate included.
In general, the radial basis function for a shell volume could be written as~\cite{Samushia2019}
\begin{equation}
    \mathcal{J}_\ell(k_{\ell n}r) \equiv j_\ell(k_{\ell n}r) + A_{\ell n} y_\ell(k_{\ell n}r) \,.
\end{equation}
where $j_\ell$ and $y_\ell$ are the spherical Bessel function of first and second kind, respectively.
The discrete wavenumbers $k_{\ell n}$ and corresponding factors $A_{\ell n}$ indexed by $n$ for each $\ell$ are determined by the Dirichlet boundary conditions.
If the field is defined in a sphere out to a certain radius, then $A_{\ell n}$ would always be zero and $\mathcal{J}_\ell$ simply reduces to $j_\ell$.
While if the field is defined in a shell with a non-zero lower radius limit, we could have non-zero $A_{\ell n}$ factors.
The galaxy samples we will consider include both of these sphere and shell cases.
With this radial eigenfunction, the SFB decomposition reads
\begin{equation} \label{eq:SFB_trans}
    f(r,\hat{r}) = \sum_{\ell m n} f_{\ell m}(k_{\ell n}) \mathcal{J}_\ell(k_{\ell n}r) Y_{\ell m}(\hat{r}) \,,
\end{equation}
with the coefficients
\begin{equation} \label{eq:SFB_coef}
\begin{split}
    f_{\ell m n} &\equiv f_{\ell m}(k_{\ell n}) \\
    &= \tau_{\ell n}^{-1} \int dr \int d\Omega\, r^2 f(r,\hat{r}) \mathcal{J}_\ell(k_{\ell n}r) Y_{\ell m}^*(\hat{r}) \,.
\end{split}
\end{equation}
The normalization factors $\tau_{\ell n}$ for different radial boundaries are derived in Appendix~\ref{sec:orthogonality}.
With Eq.~(\ref{eq:SFB_coef}), the power spectra in SFB basis can be related to the 2-point correlation function (2PCF) or the power spectra of $f(\bm{r})$ in 3D Cartesian coordinates, see more details in Appendix~\ref{sec:sfb_3d}.

\subsection{\label{subsec:power_spectra}Auto- and cross-power spectra of matter field tracers}
In what follows, we derive the SH and SFB power spectra for the 2D and 3D tracers of the matter field, whose homogeneous and isotropic power spectrum today $P_{m,0}(k)$ is defined through
\begin{equation}
    \langle\delta_{m,0}(\bm{k})\delta_{m,0}^*(\bm{k}')\rangle = (2\pi)^3 \deltaD(\bm{k}-\bm{k}') P_{m,0}(k) \,.
\end{equation}
The Fourier transform of the matter field $\delta_{m,0}(\bm{r})$ in 3D Cartesian coordinates 
\begin{equation}
    \delta_{m,0}(\bm{r}) = \int \frac{d^3 k}{(2\pi)^3} e^{i\bm{k}\cdot\bm{r}} \delta_{m,0}(\bm{k}) \,,
\end{equation}
and the plane wave expansion in spherical coordinates
\begin{equation} \label{eq:plane_wave_expansion}
    e^{i\bm{k}\cdot\bm{r}} = 4\pi \sum_{\ell m} i^{\ell} j_\ell(kr) Y_{\ell m}(\hat{r}) Y_{\ell m}^*(\hat{k}) \,,
\end{equation}
will be used.
We also assume the linear evolution of the matter field in Eq.~(\ref{eq:delta_m_evo}).

For the 2D CMB lensing and galaxy projected overdensity maps, the corresponding SH coefficients $\kappa_{\ell m}$ and $g_{\ell m}$ are given by Eq.~(\ref{eq:SH_coef}).
The angular power spectrum is defined through
\begin{equation}
    \langle a_{\ell m} a'^*_{\ell' m'}\rangle = \delta_{\ell \ell'}^{\rm K} \delta_{m m'}^{\rm K} C_\ell^{a a'} \,.
\end{equation}
With the two fields $a$ and $a'$ being either $\kappa$ or $g$, which are related to the matter field through Eq.~(\ref{eq:cmb_lensing_kappa}) and~(\ref{eq:g_model}), we can get
\begin{equation} \label{eq:cl_22}
    C_\ell^{a a'} = \frac{2}{\pi} \int dk\, k^2 P_{m,0}(k) \Delta_\ell^a(k) \Delta_\ell^{a'}(k) \,,
\end{equation}
where $\Delta_\ell^{a}$ denotes the transfer function of the 2D matter field tracer $a$.
For $\kappa$ and $g$, we have
\begin{equation}
    \Delta_\ell^\kappa(k) = \int dr\, W_\kappa(r) D(z) j_\ell(kr) \,,
\end{equation}
and
\begin{equation} \label{eq:Delta_g}
    \Delta_\ell^{g}(k) = \Delta_\ell^{g_d}(k) + \Delta_\ell^{g_r}(k) + \Delta_\ell^{g_n}(k) + \Delta_\ell^{g_m}(k) \,,
\end{equation}
which includes contributions from the main Gaussian overdensity signal ($g_d$), and also the modifications due to RSD ($g_r$), PNG ($g_n$), and magnification bias ($g_m$).
These galaxy transfer function components are
\begin{align}
    \Delta_\ell^{g_d}(k) &= \int_{r,g} b_g(z) D(z) j_\ell(kr) \,,\\
    \Delta_\ell^{g_r}(k) &= - \int_{r,g} f(z) D(z)  j_\ell''(kr) \,,\\
    \Delta_\ell^{g_n}(k) &= f_{\rm NL} \int_{r,g} \Delta b_g(z, k) D(z)  j_\ell(kr) \,,\\
    \begin{split}
    \Delta_\ell^{g_m}(k) &= (5s-2) \times \\
                           &\qquad\int_{r,g} \int_0^r dr'\, W_\kappa(r',r) D(z') j_\ell(kr') \,,
    \end{split}
\end{align}
where for simplicity we define a shorthand notation
\begin{equation}
    \int_{r,g} \equiv \int dr\,\frac{H(z)}{c} f_g(z) \,.
\end{equation}
Similar expressions have also been derived in some previous work, see e.g.~\cite{Padmanabhan2007,Slosar2008,Yang2018}.

As shown above in Eq.~(\ref{eq:SFB_trans}), 3D SFB transform is a natural extension to the 2D SH transform in Eq.~(\ref{eq:SH_trans}), with the same angular eigenfunctions indexed by $\ell m$.
Thus SH coefficients $a_{\ell m}$ in Eq.~(\ref{eq:SH_coef}) can be cross-correlated with SFB coefficients $f_{\ell m n}$ in Eq.~(\ref{eq:SFB_coef}) of any radial mode indexed by $n$
\begin{equation}
    \langle a_{\ell' m'} f_{\ell m n}^* \rangle = \delta_{\ell \ell'}^{\rm K} \delta_{m m'}^{\rm K} C_{\ell n}^{a f} \,,
\end{equation}
which gives
\begin{equation} \label{eq:cl_23}
    C_{\ell n}^{a f} = \frac{2}{\pi} \int dk\, k^2 P_{m,0}(k) \Delta_\ell^a(k) \Delta_{\ell n}^f(k) \,,
\end{equation}
where $\Delta_{\ell n}^f(k)$ is the transfer function of $f(r,\hat{r})$, a 3D tracer of the matter field like $\delta$ in this work.
Similarly, the correlation between two 3D fields in SFB basis reads,
\begin{equation}
    \langle f_{\ell m n}f'^*_{\ell' m' n'}\rangle = \delta_{\ell \ell'}^{\rm K} \delta_{m m'}^{\rm K} C_{\ell n n'}^{f f'} \,,
\end{equation}
with
\begin{equation} \label{eq:cl_33}
    C_{\ell n n'}^{f f'} = \frac{2}{\pi} \int dk\, k^2 P_{m,0}(k) \Delta_{\ell n}^f(k) \Delta_{\ell n'}^{f'}(k) \,.
\end{equation}
where in general the radial modes are not orthonormal due to the radial selection and evolution of the fields.
In this case, for each $\ell$ the power spectrum is a covariance matrix of the radial modes.
Similar as that for the projected galaxy map $g(\hat{r})$, for the 3D overdensity field $\delta(r,\hat{r})$ in Eq.~(\ref{eq:delta_model}), the transfer function is given as
\begin{equation} \label{eq:Delta_delta}
    \Delta_{\ell n}^{\delta}(k) = \Delta_{\ell n}^{\delta_d}(k) + \Delta_{\ell n}^{\delta_r}(k) + \Delta_{\ell n}^{\delta_n}(k) + \Delta_{\ell n}^{\delta_m}(k) \,,
\end{equation}
with
\begin{align}
    \Delta_{\ell n}^{\delta_d}(k) &= \int_{r,\delta} b_g(z) D(z) j_\ell(kr) \,,\\
    \Delta_{\ell n}^{\delta_r}(k) &= - \int_{r,\delta} f(z) D(z) j_\ell''(kr) \,,\\
    \Delta_{\ell n}^{\delta_n}(k) &= f_{\rm NL} \int_{r,\delta} \Delta b_g(z, k) D(z) j_\ell(kr) \,,\\
    \begin{split}
    \Delta_{\ell n}^{\delta_m}(k) &= (5s-2)\times \\
                                  &\qquad \int_{r,\delta}\int_0^r dr'\, W_\kappa(r',r) D(z') j_\ell(kr') \,,
    \end{split}
\end{align}
where the shorthand notation
\begin{equation}
    \int_{r,\delta} \equiv \tau_{\ell n}^{-1} \int dr\, r^2 \phi_g(r) \mathcal{J}_\ell(k_{\ell n}r) \,.
\end{equation}
Notice that the $\phi_g(r)-1$ term in Eq.~(\ref{eq:delta_model}) is independent of the angular direction and thus only contributes to the monopole ($\ell=0$), which along with the dipole ($\ell=1$) will not be included in the Fisher analyses in this work.

The numerical computation of the transfer functions requires the line-of-sight integrals over the highly oscillatory spherical Bessel functions $j_\ell(kr)$ and $y_\ell(kr)$, for which we include more details in Appendix~\ref{sec:num_cal}.
In this work, we use \textsc{Colossus}~\cite{Diemer2018} and \textsc{CAMB}~\cite{web-camb,Lewis2000} to calculate the required cosmological functions, including the 3D matter power spectrum.

\subsection{\label{subsec:noise}Noise in auto-power spectra}
In this work, we consider both auto- and cross-power spectra.
Usually the noise in one observable is not correlated with signal and noise in another different observable, thus being independent of noise is one advantage of cross-correlation.
Below we consider the well-known noise expressions in the auto-power spectra, including the lensing reconstruction noise for CMB lensing and the shot noise for galaxy overdensity.

\subsubsection{\label{subsec:lensing_noise}CMB lensing reconstruction noise}
\begin{figure}
    \centering
    \includegraphics[width=\columnwidth]{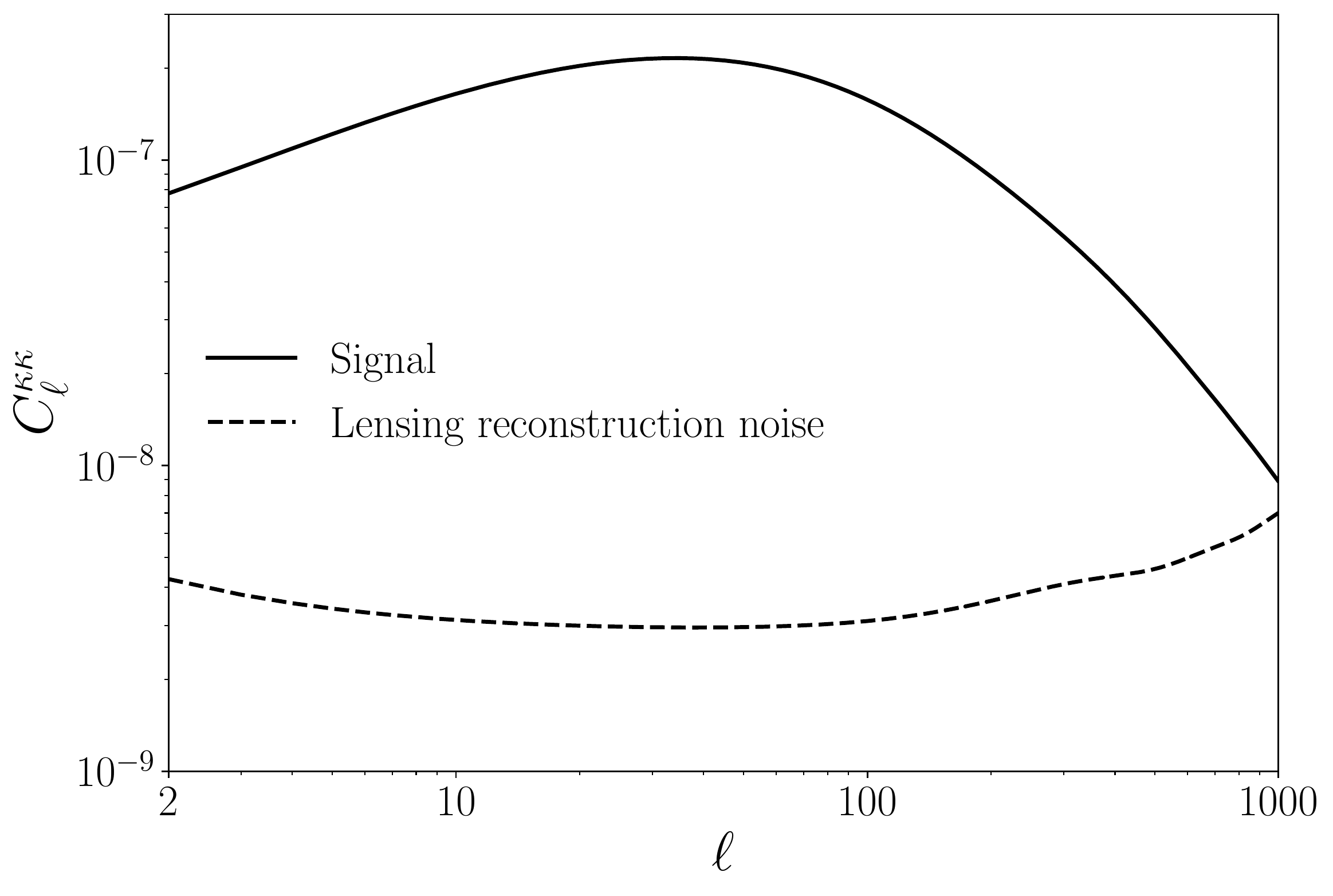}
    \caption{\label{fig:clkk}CMB lensing convergence power spectrum and the lensing reconstruction noise discussed in Section~\ref{subsec:lensing_noise}.}
\end{figure}

Weak lensing of the CMB induces correlations between different Fourier modes of the CMB temperature and polarization fields.
As a result, estimators of $\kappa$ field out of linear combinations of terms quadratic in different modes of observed temperature and polarization fields can be constructed \cite{Hu_01}.
Indeed, almost all the CMB lensing analyses to date have relied upon such quadratic estimators.
Recently, \cite{Maniyar2021} showed that the well-known Hu and Okamoto \cite{Hu_02} estimator is not the most optimal quadratic estimator that can be constructed out of the temperature and polarization maps as was previously thought.
They instead derive the global-minimum-variance (GMV) estimator built out of all possible quadratic combinations of T, E, and B (temperature, and E and B mode polarization).
Here, we use the GMV estimator to estimate the noise of the reconstructed $C_\ell^{\kappa \kappa}$ denoted by $N_\ell^{\kappa \kappa}$.
\begin{eqnarray}
N_\ell^{\kappa \kappa} &=& 2 \int \Xi_{ij}(\bl_1, \bl_2) \Xi_{pq}(\bl_1, \bl_2) C_{l_1}^{ip} C_{l_2}^{jq}\, ,
\label{eq:variance_gmv}
\end{eqnarray}
where $\Xi_{ij}(\bl, \bl')$ is a three by three symmetric matrix corresponding to weights applied to the T, E, and B mode pairs and is derived in \cite{Maniyar2021}, and $C_{l}^{ij}$ are CMB power spectra.
The CMB lensing signal and the reconstruction noise (with survey specifications in Section~\ref{subsec:cmb_lensing_survey}) are shown in Fig.~\ref{fig:clkk}.
\cite{Schmittfull2018} divide the noise coming from the `EB' estimator by a factor of 2.5, to approximately match the noise level expected by the iterative reconstruction process.
We find that this results in an overall noise reduction by a factor of $\sim 2$ for the minimum variance coming out of the Hu and Okamoto estimator.
Here we assume that a similar reduction of a factor of two will take place for the GMV estimator as well and thus divide the minimum variance noise from the GMV estimator by two as shown in Fig.~\ref{fig:clkk}.

\subsubsection{Galaxy shot noise}
Galaxy shot noise is induced by the discrete nature of the point targets. Assuming Poisson sampling~\cite{Feldman1994}, the shot noise $\epsilon^n(\bm{r})$ contribution to the 2PCF of the number density field $n(r,\hat{r})$ is shown~\cite{Yoo2013} to be
\begin{equation}
    \langle\epsilon^n(\bm{r})\epsilon^n(\bm{r}')\rangle = \bar{n}_V\phi_g(r)\deltaD(\bm{r}-\bm{r}') \,,
\end{equation}
from which the shot noise power spectra in spherical basis can be derived for the 2D and 3D overdensity fields starting from the definitions of SH and SFB coefficients.

For the 2D overdensity field $g(\hat{r})$ defined in terms of $n(r,\hat{r})$ in Eq.~(\ref{eq:g_data}), the corresponding angular shot noise is $\epsilon^g(\hat{r})\equiv \int dr\,r^2\epsilon^n(\bm{r})/\bar{n}_\Omega$.
The shot noise power spectrum is given by
\begin{equation}
    \langle \epsilon^g_{\ell m}\epsilon^{g*}_{\ell'm'} \rangle = \deltaK_{\ell\ell'}\deltaK_{m m'} N^{gg}_{\ell} \,,
\end{equation}
with
\begin{equation} \label{eq:shot_noise_SH}
    N^{gg}_{\ell} = \frac{\bar{n}_V}{\bar{n}^2_\Omega}\int dr\, r^2\phi_g(r) = \frac{1}{\bar{n}_\Omega} \int dz f_g(z) = \frac{1}{\bar{n}_\Omega} \,,
\end{equation}
which is simply the inverse of the average angular number density that is a constant for all the angular modes and independent of the redshift distribution.

Similarly, for the 3D overdensity field $\delta(r,\hat{r})$ defined in Eq.~(\ref{eq:delta_data}), the shot noise field $\epsilon^\delta(\bm{r})\equiv \epsilon^n(\bm{r})/\bar{n}_V$ has the power spectrum
\begin{equation}
    \langle\epsilon^\delta_{\ell m n}\epsilon^{\delta*}_{\ell' m' n'}\rangle = \deltaK_{\ell\ell'}\deltaK_{m m'} N^{\delta\delta}_{\ell n n'} \,,
\end{equation}
with
\begin{equation} \label{eq:shot_noise_SFB}
    N^{\delta\delta}_{\ell n n'} = \frac{\tau_{\ell n}^{-1} \tau_{\ell n'}^{-1}}{\bar{n}_V} \int dr\, r^2\phi_g(r)\mathcal{J}_\ell(k_{\ell n}r)\mathcal{J}_\ell(k_{\ell n'}r) \,,
\end{equation}
where in general we could have non-zero shot noise for the cross correlation between different radial modes.
For top-hat $\phi_g(r)$ (i.e. $\phi_g(r)=1$ in the survey coverage, otherwise $0$), the orthogonality relation (see Appendix~\ref{sec:orthogonality}) could be used and the RHS of Eq.~\ref{eq:shot_noise_SFB} reduces to $\deltaK_{n n'}\tau_{\ell n}^{-1}/\bar{n}_V$.
However, this is usually not the case for real galaxy surveys.

\section{\label{sec:surveys}Fiducial surveys}
In this section, we describe the fiducial survey setups for the Fisher forecasts.

\subsection{\label{subsec:cmb_lensing_survey}CMB lensing survey}
For the CMB lensing survey, we consider the CMB-S4~\cite{CMB-S4-science2016} level precision with the white noise of the detector given by $\Delta_T=1\,\mu K\,{\rm arcmin}$ and $\Delta_P=\sqrt{2}\Delta_T$.
The lensing reconstruction noise level is shown in Fig.~\ref{fig:clkk}.

\subsection{\label{subsec:galaxy_surveys}Galaxy redshift surveys}
\begin{table*}
    \caption{\label{tab:galaxy_surveys}Specifications of the galaxy survey samples considered, including the total number of targets $N_g$, fractional sky coverage $f_{\rm sky}$, redshift coverage $z$, radial comoving width $r_{\rm width}$, average comoving volume density $\bar{n}_V$, and redshift uncertainty $\tilde{\sigma}_z \equiv \sigma_z/(1+z)$.
    More details are discussed in Section~\ref{subsec:galaxy_surveys}.}
    \begin{ruledtabular}
    \begin{tabular}{lddlrdd}
        Sample    & \multicolumn{1}{c}{$N_g$} & \multicolumn{1}{c}{$f_{\rm sky}$} & \multicolumn{1}{c}{$z$} & \multicolumn{1}{c}{$r_{\rm width}$} &
        \multicolumn{1}{c}{$\bar{n}_V$} & \multicolumn{1}{c}{$\tilde{\sigma}_z$} \\
         & \multicolumn{1}{c}{[M]} & \multicolumn{1}{c}{[\%]} & & \multicolumn{1}{c}{[Mpc]} &
         \multicolumn{1}{c}{[$\times 10^{-3}\,{\rm Mpc}^{-3}$]} & \\
        \hline
        DESI BGS  & 9.8  & 33.9 & 0 - 0.5    & 1954 & 0.92 & $-$ \\
        DESI ELG  & 17   & 33.9 & 0.6 - 1.7  & 2561 & 0.12 & $-$ \\
        \textit{Euclid} & 61   & 36.4 & 0.6 - 2.1  & 3177 & 0.26 & $-$ \\
        LSST low-$z$ & 2801 & 48.5 & 0 - 2 & 5314 & 9.2 & 0.05 \\
        LSST high-$z$ & 865 & 48.5 & 2 - 5 & 2631 & 1.2 & 0.05 \\
        SPHEREx 1 & 24 & 75 & 0 - 1.4 & 4292 & 0.096 & 0.003 \\
        SPHEREx 2 & 76 & 75 & 0 - 1.4 & 4292 & 0.31 & 0.01 \\
        SPHEREx 3 & 147 & 75 & 0 - 1.4 & 4292 & 0.59 & 0.03 \\
    \end{tabular}
    \end{ruledtabular}
\end{table*}

\begin{figure}
    \centering
    \includegraphics[width=\columnwidth]{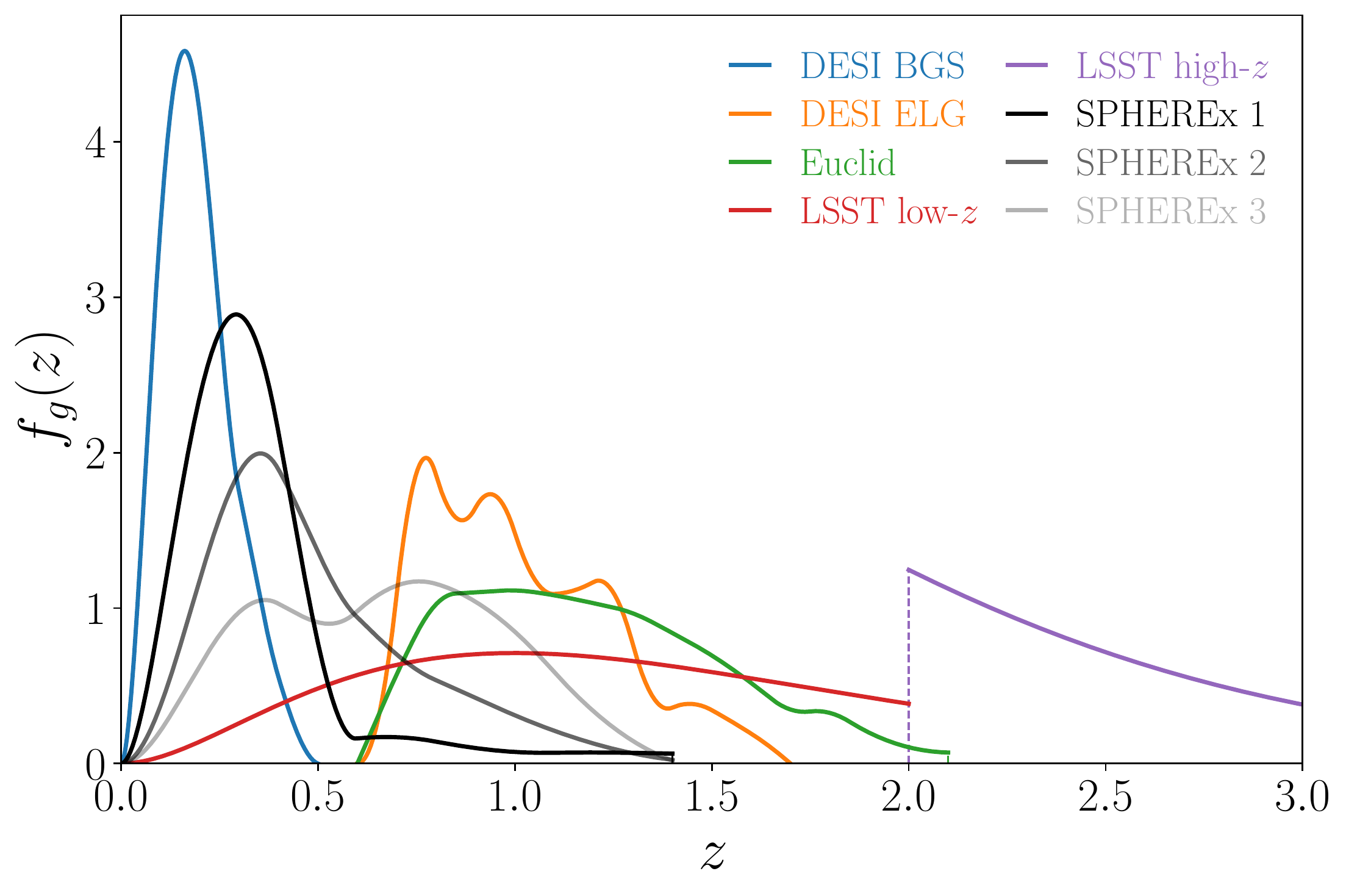}
    \caption{\label{fig:fg}Redshift distributions of the galaxy samples, the integrals of which are normalized to one in the redshift range covered. Note that the LSST high-$z$ sample extends to $z=5$ though we truncate the plot at $z=3$ for better clarity.}
\end{figure}

\begin{figure}
    \centering
    \includegraphics[width=\columnwidth]{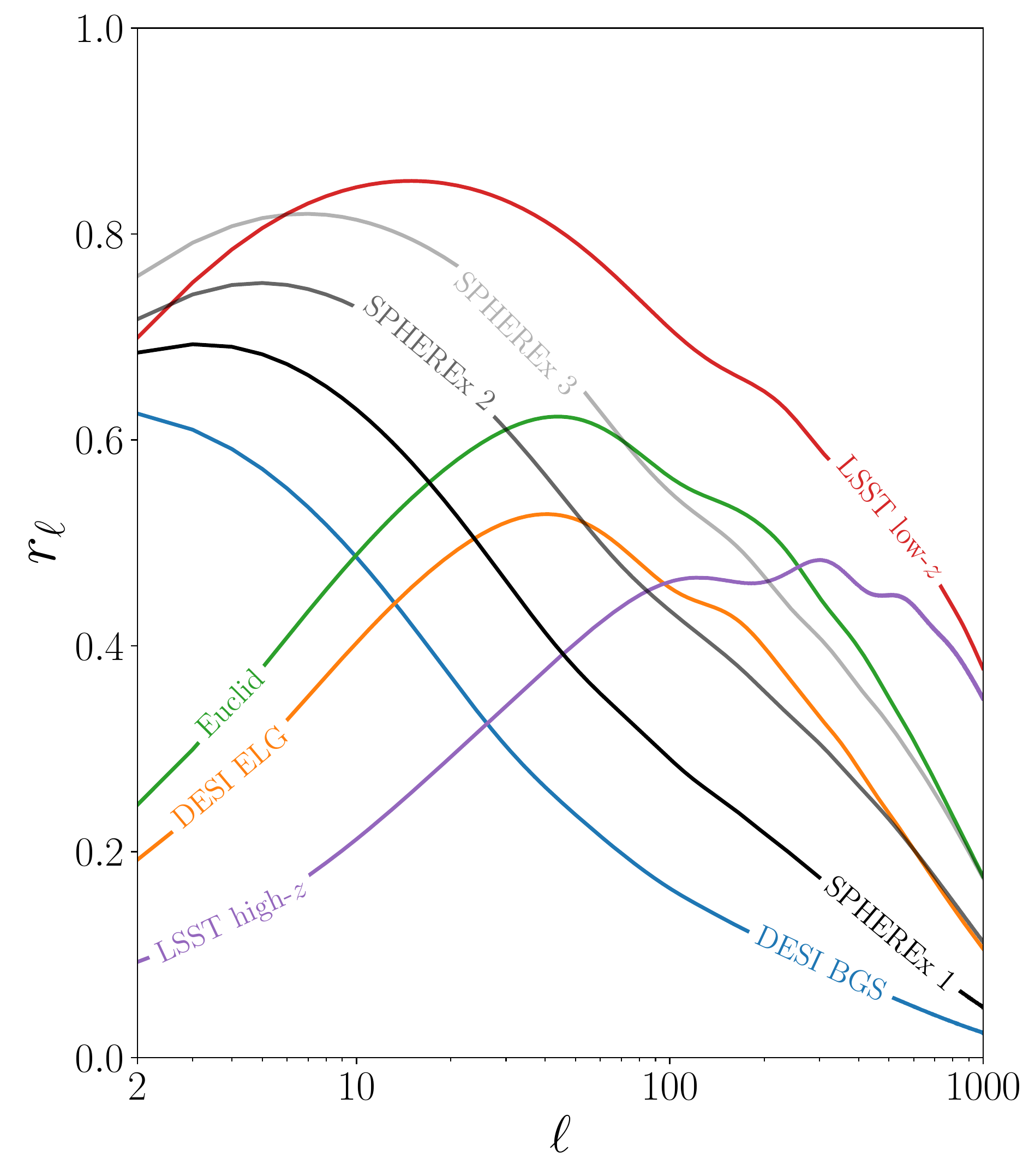}
    \caption{\label{fig:clkg_corr_coef}Cross-correlation coefficients of CMB lensing and the galaxy samples, which are calculated with Eq.~(\ref{eq:cross_corr_coef}) by projecting each galaxy sample into a 2D angular map.}
\end{figure}

We consider a few fiducial galaxy samples that mimic the designed specifications of some future spectroscopic surveys.
Although we simply use the name of the surveys to denote the samples in this work and omit the ``-like'' suffix for brevity, it is worth being reminded that the real data from these surveys could be more or less different.
In our analysis, each survey or its sub-samples can be completely described with the total number of targets $N_g$, the fractional sky coverage $f_{\rm sky}$, the redshift distribution $f_g(z)$, and a redshift-dependent clustering bias model $b_g(z)$.
Some details of the galaxy samples below are summarized in Table~\ref{tab:galaxy_surveys}, and the redshift distributions are shown in Fig.~\ref{fig:fg}.

The spectroscopic surveys we consider include DESI~\cite{web-desi} and the redshift survey of the \textit{Euclid} satellite mission~\cite{web-euclid}.
For DESI, we include the two largest sub-samples, the bright galaxy sample (BGS) and the emission line galaxy (ELG) sample.
The expected redshift distributions are given by Table 2.3 and 2.5 in~\cite{DESI-science2016}, and we assume the fiducial redshift-dependent bias $b_g(z)|_{\rm BGS} = 1.34 / D(z)$~\cite{DESI-science2016} and $b_g(z)|_{\rm ELG} = 0.84 / D(z)$~\cite{Mostek2013}.
For Euclid~\cite{Euclid2011,Euclid2020}, number densities are taken from Table 3 in~\cite{Euclid2018} and we assume the reference case (i.e. the $n_2$ column).
We take a fiducial bias $b_g(z)|_{\rm Euclid}=0.76/D(z)$~\cite{Font-Ribera2014}.
The galaxy redshifts in the spectroscopic surveys are measured to very high accuracy, whose uncertainties correspond to scales that are much smaller than the scales we consider in this work.
Thus for these samples, we shall just ignore the redshift uncertainty.

Besides, we also consider one photometric and another spectro-photometric survey which could have non-negligible redshift uncertainties.
The photometric one is the LSST~\cite{web-lsst} survey, which is expected to have a redshift distribution given by~\cite{LSST2009}
\begin{equation}
\begin{split}
        \frac{d^2N}{dzd\Omega} =\ &\frac{\bar{n}^{\rm LSST}_{\Omega,{\rm tot}}\beta}{z_*\Gamma\left[(\alpha+1)/\beta\right]} \\
        &\times\left(\frac{z}{z_*}\right)^\alpha \exp\left[-\left(\frac{z}{z_*}\right)^\beta\right] \ \ \deg^{-2} \,,
\end{split}
\end{equation}
with $\alpha=2.0$, $\beta=1.0$, $z_*=0.5$, and a total projected number density $\bar{n}^{\rm LSST}_{\Omega,{\rm tot}}=50\,{\rm arcmin}^{-2}$.
We consider the redshift depth up to $z=5$, which encloses more than $99.7 \%$ of the total targets.
The bias model is assumed to be $b_g(z)|_{\rm LSST} = 0.95/D(z)$~\cite{Font-Ribera2014}.
The LSST redshift coverage is really wide and in this work, we divide the LSST sample into two sub-samples, with one covering lower redshift $z=0-2$ and the other covering higher redshift $z=2-5$.
The spectro-photometric one is the SPHEREx~\cite{web-spherex} survey.
SPHEREx provides forecasts of galaxy number density and bias for five sub-samples based on the redshift uncertainty~\cite{SPHEREx2014}.
We use the three samples with $\tilde{\sigma}_z\leq 0.03$, denoted as SPHEREx \{1, 2, 3\}.
The number density distribution and bias functions are interpolated from data in this public products repository~\cite{web-spherex-products}.

Since we are doing joint analyses of these galaxy samples with CMB lensing, before doing Fisher forecasts we can do a quick check on the cross-correlation strength, which can be quantified with coefficients
\begin{equation} \label{eq:cross_corr_coef}
    r_\ell = \frac{C_\ell^{\kappa g}}{[(C_\ell^{\kappa\kappa}+N_\ell^{\kappa\kappa})(C_\ell^{gg}+N_\ell^{gg})]^{1/2}} \,,
\end{equation}
where $g$ is the projected overdensity map of the galaxy sample over its whole redshift coverage, and noises are added to the corresponding auto-power spectra.
These cross-correlation coefficients are shown in Fig.~\ref{fig:clkg_corr_coef}.
In general, the coefficients peak at different angular scales with galaxy samples covering lower redshifts peaking at larger angular scales and vice-versa.
With the same CMB lensing reconstruction noise, the overall amplitude is mainly determined by the galaxy shot noise and also the redshift overlap with the CMB lensing kernel that peaks around $z=2$.

\section{\label{sec:fisher}Fisher forecast setup}
In this section, we discuss the setups for the Fisher forecast on parameters of interest with the joint analysis of CMB lensing and galaxy overdensity fields, for which both SFB and TSH analyses will be considered for a comparison.

\subsection{\label{subsec:parameters}Parameters}
The parameters of primary interest are the PNG amplitude $f_{\rm NL}$ and the RSD exponent $\gamma$.
We assume fiducial values $f_{\rm NL}=0$, i.e. no PNG and $\gamma=0.55$, the GR prediction.

We also include a few free nuisance parameters that account for the uncertainties in some galaxy properties.
For the fiducial clustering bias model, we assume the redshift-dependence is well known in the redshift range covered while introducing a constant parameter $A_b$, which is free for tuning the overall amplitude around the fiducial value $A_b=1$.
Another free parameter is the foreground magnification bias $s$, for which we take a fiducial value $s=0.4$ assuming no distortion, see Eq.~(\ref{eq:mag_bias}).
It is important to notice that unlike $f_{\rm NL}$, $\gamma$ or other cosmological parameters, $A_b$ and $s$ depend on the particular galaxy sample.
The derivatives of power spectra from one sample with respect to these two parameters of another sample would simply be zero.
For example, if we have two galaxy samples in the joint analysis, then besides other parameters, the free parameter set will include $\{A_b^1,\,A_b^2,\,s^1,\,s^2\}$ with 1 and 2 denoting the two samples.
The power spectra of sample 1 should be independent of $A_b^2$ and $s^2$.

The parameters above are all associated with the galaxy field, and they appear only in the galaxy transfer functions (Eq.~(\ref{eq:Delta_g}) and~(\ref{eq:Delta_delta})).
In addition, we also consider the dependence of the matter field, i.e. its power spectra today $P_{m,0}(k)$ and linear growth factor $D(z)$, on the background cosmological parameters $\{H_0,\,\Omega_{m,0},\,\Omega_{b,0},\,\sigma_8,\,n_s\}$, which are also included in the Fisher analyses.
We assume a flat $\Lambda$CDM cosmology with \textit{Planck} 2018 CMB TT,TE,EE+lowE best-fit results~\cite{Planck2018-cosmo} as fiducial values.
It would also be helpful to include the \textit{Planck} constraints as prior information in our Fisher analyses, with more details in Section~\ref{subsec:prior_info}.
Notice that CMB lensing depends on these background cosmological parameters but not $f_{\rm NL}$ or $\gamma$, which can help reducing the degeneracy, and this is one of the motivations for the joint analyses.

\subsection{Linear SFB and TSH modes} \label{subsec:linear_modes}
Some of the theoretical or fiducial models discussed above, e.g. the linear evolution of the matter field, scale-independent galaxy clustering bias, and the RSD correction etc., are valid only on large linear scales.
In what follows, we discuss the linear SFB and TSH modes that will be included in our Fisher analyses.

There have been several ways of quantifying the 3D threshold $k_{\rm max}^{\rm 3D}(z)$ between linear (or quasi-linear) and nonlinear Fourier modes based on the linear matter power spectrum $P_m(k,z)$.
For example, we may simply set a limit to the dimensionless matter power spectrum $\Delta^2_m(k,z) \equiv k^3P_m(k,z)/(2\pi^2)$, whose value is monotonically increasing with $k$.
Some previous work defines the linear scales as those satisfying $\Delta^2_m(k,z) < 1$.
Another slightly more complicated way that has been widely used is to evaluate the variance of the smoothed matter field $\sigma^2(R,z)=\int \frac{d^3k}{(2\pi)^3}\, W^2(kR) P_m(k,z)$, where $W(kR)= 3[\sin(kR)-kR\cos(kR)]/(kR)^3$ for a tophat filter function in real space~\cite{Pierpaoli2001}.
This variance is decreasing with $R$ and by requiring $\sigma^2(R,z) < 1$, we could get the minimum radius $R_{\rm min}$ and the corresponding $k^{\rm 3D}_{\rm max}=1/R_{\rm min}$.
One more criterion is based on the impact of the nonlinear correction (e.g. with a halofit model in~\cite{Mead2016}) to linear $P_m(k,z)$.
The threshold can be quantified by requiring the fractional impact of the correction to be within e.g. $10\,\%$.

All these three methods above can be used to determine $k_{\rm max}^{\rm 3D}(z)$.
Then a very natural idea is to convert this threshold on the 3D wavenumber to the limit on SFB and TSH modes.
However, these conversions are not clearly defined for a few reasons.
First, power spectra of SFB and TSH modes are given by the integral over 3D wavenumbers, as shown in Eq.~(\ref{eq:cl_22}) and~(\ref{eq:cl_33}).
The contributions of different 3D wavenumbers to these integrals depend on the boundary condition for SFB (see Appendix~\ref{sec:sfb_3d}) and the bin size for TSH, and also the redshift-dependent functions.
Even just for a single bin with SH analysis, Limber approximation which picks out a particular $k \simeq \ell/r$ only works for high $\ell$s and wide bins, which is not always satisfied in our TSH analyses.
On the other hand, for TSH analysis, besides the $\ell$ for each bin, which is usually approximated with $k r$, we still need to determine the bin size.
However, we find it hard to properly determine a pair of the $\ell_{\rm max}$ value and the tomographic bin size that corresponds to a given $k_{\rm max}^{\rm 3D}(z)$.

Given these issues and inspired by the third method for determining $k_{\rm max}^{\rm 3D}(z)$ above, we quantify the linear modes for SFB and TSH analyses independently.
We evaluate Eq.~(\ref{eq:cl_22}) and~(\ref{eq:cl_33}) using a linear and non-linear matter power spectrum $P_{m}(k,z)$ for TSH and SFB analysis respectively.
The highest possible value of $\ell_{\rm max}$ for a given $z$ and redshift bin width in the TSH analysis (or $k_{\ell n}$ for a given $\ell$ for the SFB analysis) is then determined such that the fractional difference in the evaluation of the power spectrum using a linear and non-linear matter power spectrum is less than $10\,\%$. These are the linear modes used for the Fisher analysis.
In the meantime, the lowest $k_{\ell n}$ available for each $\ell$ for the SFB analysis are determined by the radial coverage (i.e. shell or sphere) and the Dirichlet boundary condition.
With their linear modes being determined independently, it is not guaranteed that the modes corresponding to the same scales are included for SFB and TSH analyses, which is hard to do since they behave differently in mixing 3D wavenumbers.
To better illustrate our linear modes selection, as an example, here we show the diagrams for the \textit{Euclid} sample.
\begin{figure}
    \centering
    \includegraphics[width=\columnwidth]{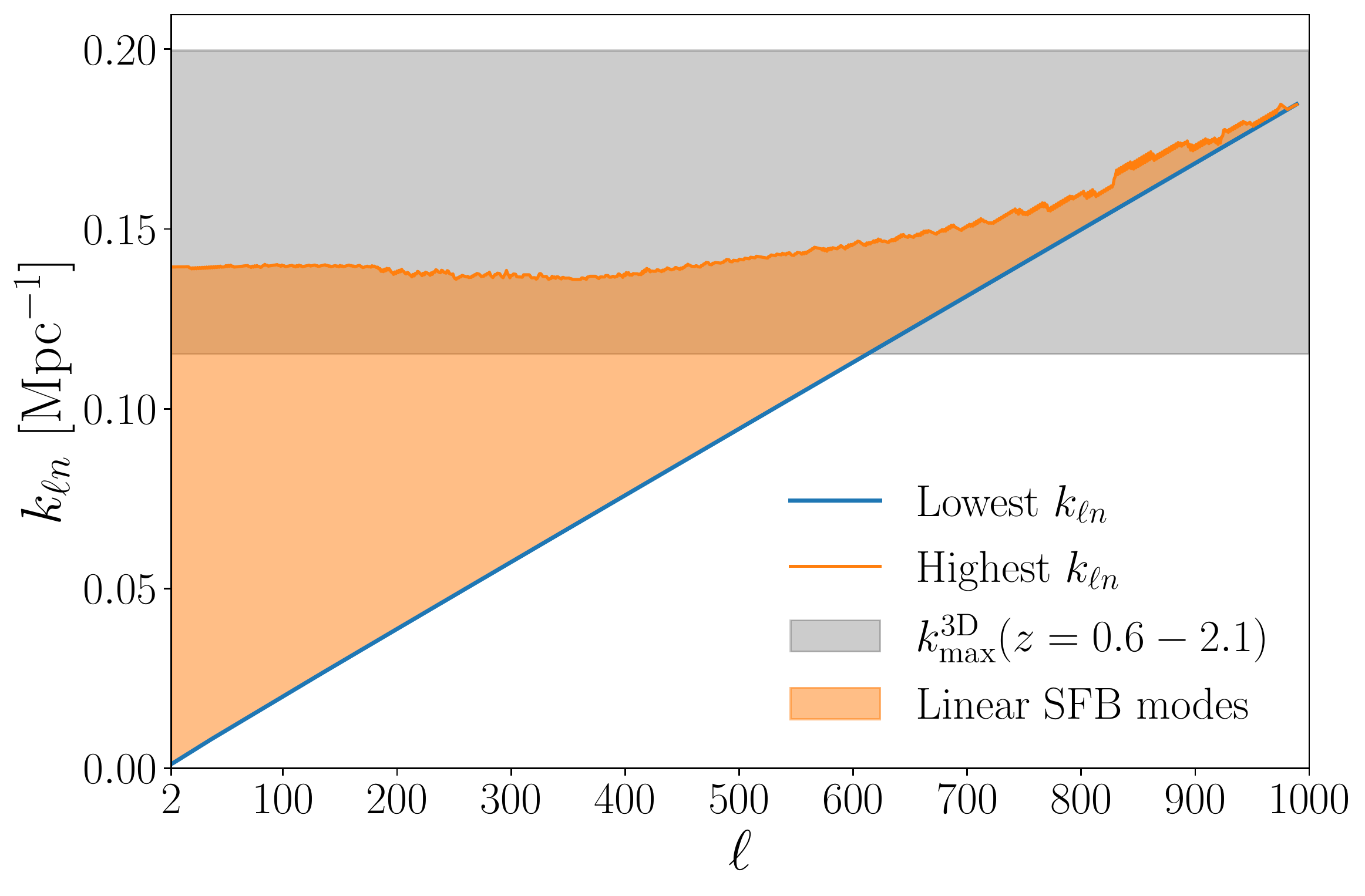}
    \caption{\label{fig:sfb_linear_modes_euclid}Linear modes in SFB analysis of the \textit{Euclid} galaxy sample, where each mode is specified by the angular multipole $\ell$ and the discrete radial wavenumber $k_{\ell n}$.
    The blue line show the lowest $k_{\ell n}$ we could have for each $\ell$ under the boundary condition, while the orange line denote the highest $k_{\ell n}$ determined by the linear requirement.
    Thus all the linear modes that will be included in the Fisher analysis are covered by the enclosed orange area.
    As a reference, we also show the maximum 3D linear wavenumber $k_{\rm max}^{\rm 3D}$, which evolves with redshift, in the grey shaded area.
    See Section~\ref{subsec:linear_modes} for more details.}
\end{figure}
In Fig.~\ref{fig:sfb_linear_modes_euclid}, the linear SFB modes are shown in the $(k_{\ell n},\,\ell)$ space.
The maximum $k_{\ell n}$ values lie between $k_{\rm max}^{\rm 3D}$ at redshift $z=0.6$ and $z=2.1$.
\begin{figure}
    \centering
    \includegraphics[width=\columnwidth]{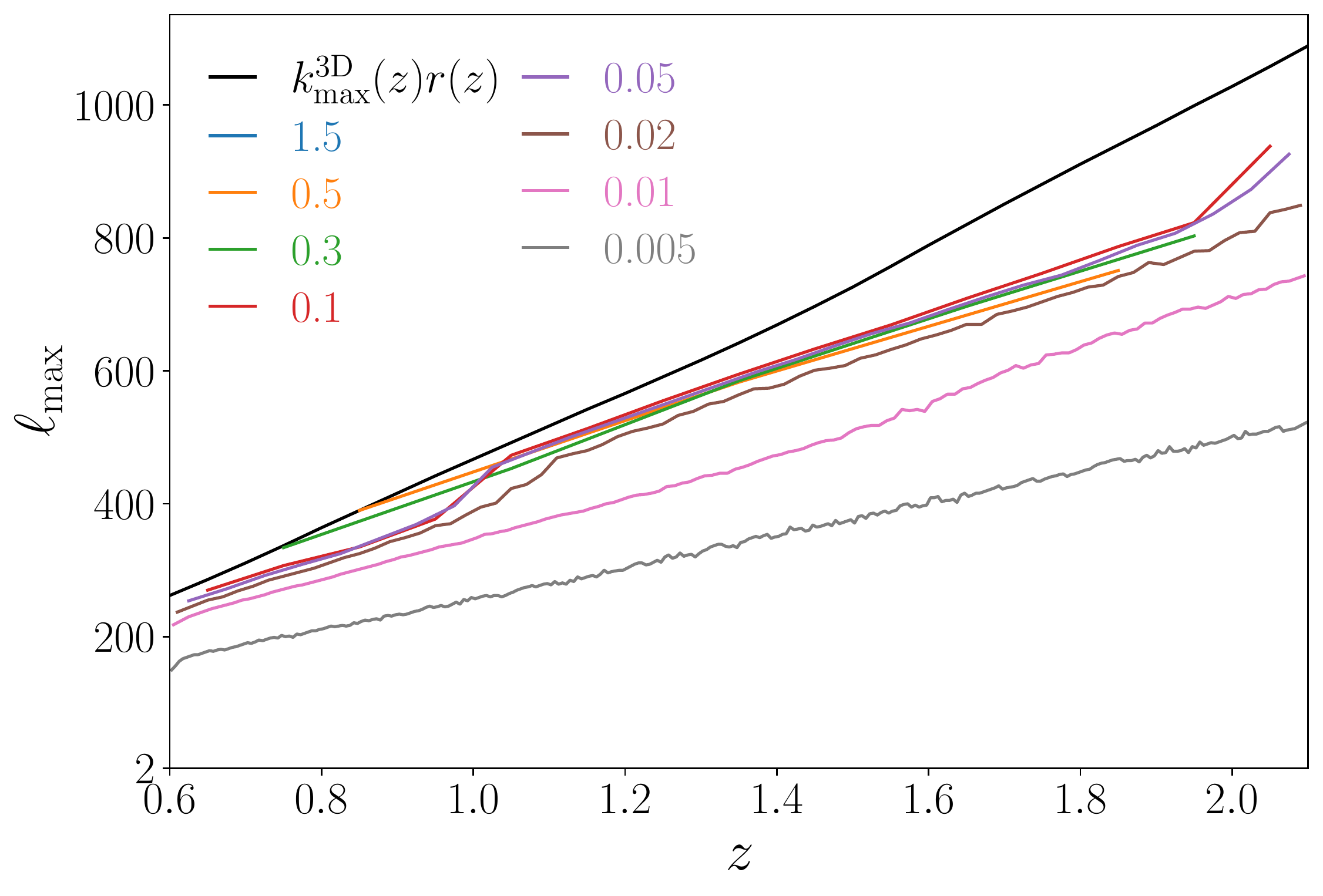}
    \caption{\label{fig:tsh_linear_lmax_euclid}Linear modes in TSH analysis of the \textit{Euclid} galaxy sample, where each mode is specified by the redshift bin size and the angular multipole $\ell$.
    Each line labelled by the tomographic redshift bin size shows the maximum $\ell$ for each bin determined by the linear requirement.
    As a reference, we also show $k_{\rm max}^{\rm 3D}(z)r(z)$, which is usually used to convert $k$ to $\ell$ for a single redshift bin.
    See Section~\ref{subsec:linear_modes} for more details.}
\end{figure}
Similarly, we have Fig.~\ref{fig:tsh_linear_lmax_euclid} for the TSH modes.
We can see that as the bin size becomes much smaller, e.g. from $0.01$ to $0.005$, we have more bins but the linear $\ell_{\rm max}$ value for each bin becomes significantly lower.
This is expected since with smaller bin sizes, more nonlinear scales are being mixed in the integral and high $\ell$ modes that become more nonlinear are excluded.
Then it is interesting to check how the total information, e.g. in terms of the signal-to-noise ratio (SNR) or constraints on different parameters, will change accordingly.
It should be expected that with only the linear modes being selected, the information would stop increasing as the bin size is decreased to a certain level.
As will be discussed in Section~\ref{subsec:TSH_bin_size} and~\ref{subsec:constraints_on_parameters}, we find that for spectroscopic samples, TSH analysis with a bin size around $\Delta z=0.01$ gives the highest overall SNR for power spectra and the tightest constraints on parameters.

\begin{figure}
    \centering
    \includegraphics[width=\columnwidth]{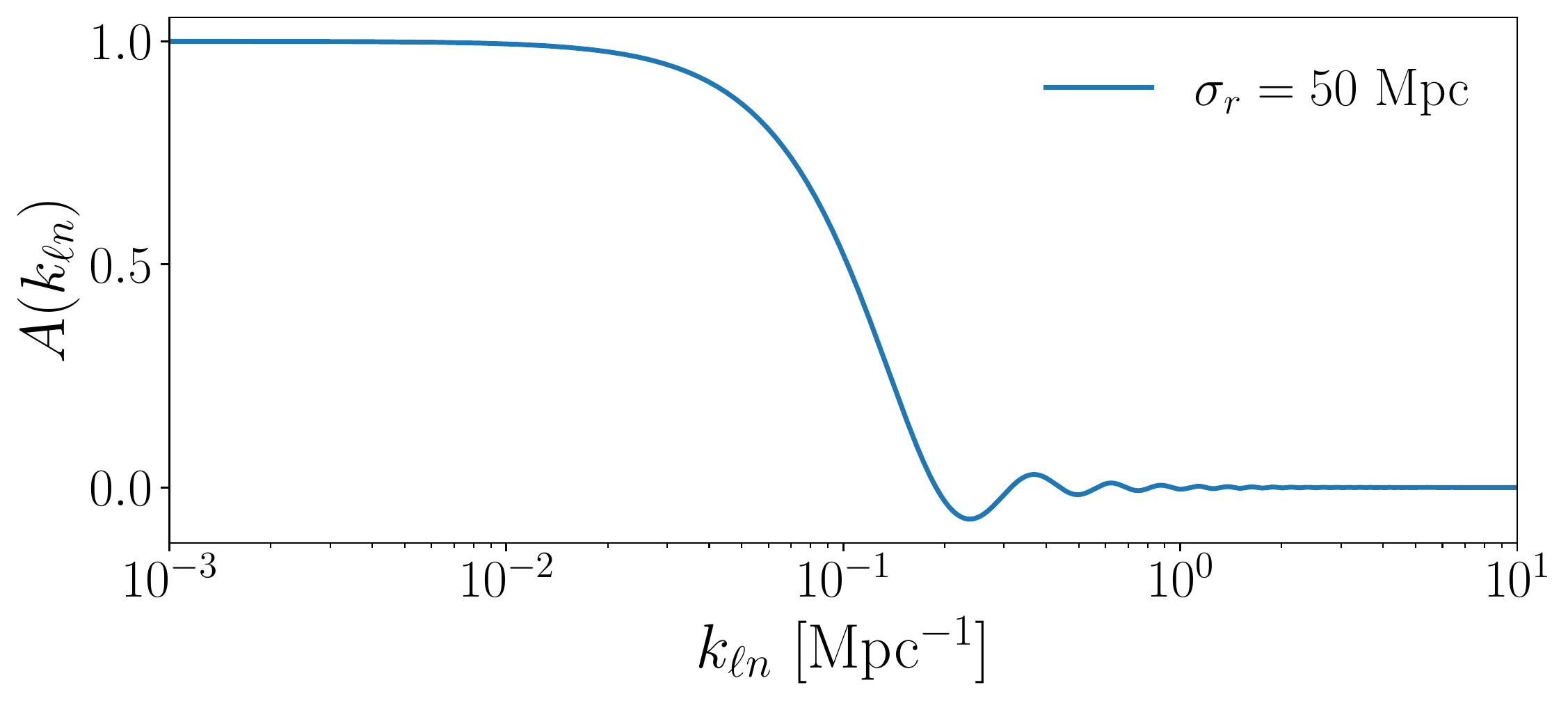}
    \caption{\label{fig:spec_window}Spectral window function (Eq.~(\ref{eq:spec_window})) multiplied to the galaxy SFB modes. The example line is shown for a redshift uncertainty $\sigma_r=50$ Mpc, which roughly corresponds to $\tilde{\sigma}_z=0.01$ centered at $z=0.7$ (e.g. the SPHEREx 2 sample).}
\end{figure}
For (spectro-)photometric samples with non-negligible redshift uncertainties, a spectral window function~\cite{Chakraborty2019}
\begin{equation} \label{eq:spec_window}
    A(k_{\ell n}) = \frac{\pi}{4}\left[ \text{sinc}\left(\frac{k_{\ell n}\sigma_r-\pi}{2}\right) + \text{sinc}\left(\frac{k_{\ell n}\sigma_r+\pi}{2}\right) \right]
\end{equation}
is applied to the galaxy transfer functions (Eq.~(\ref{eq:Delta_delta})) in SFB analyses, where $\sigma_r = c\,\tilde{\sigma}_z(1+z)/H(z)$ is the radial interval that corresponds to the redshift uncertainty.
As an example, we show $A(k_{\ell n})$ with $\sigma_r=50$ Mpc in Fig.~\ref{fig:spec_window}.
In TSH analyses, we use bin sizes that are larger than the redshift uncertainty.

\subsection{Gaussian likelihood and Fisher matrix}
The likelihood function $\mathcal{L}(\datavh|\bm{\theta})$ describes the probability of having the observed data vector $\datavh$ given the set of parameters $\bm{\theta}$, or vice versa.
For brevity, we adopt a frequently used shorthand notation for the partial derivative with respect to parameters, $\Box_{,\alpha}\equiv\partial \Box/\partial \theta_\alpha$.
The elements of Fisher matrix are defined as
\begin{equation}
    F_{\alpha\beta} \equiv \langle-(\ln\mathcal{L})_{,\alpha\beta}\rangle \,,
\end{equation}
whose inverse gives the Gaussian covariance matrix of the parameters
\begin{equation}
    {\rm Cov}(\theta_\alpha, \theta_\beta) = \left(F^{-1}\right)_{\alpha\beta} \,.
\end{equation}
Then the uncertainty of a parameter with all the other parameters being marginalized is simply given by the diagonal elements
\begin{equation}
    \sigma(\theta_\alpha) = \sqrt{\left(F^{-1}\right)_{\alpha\alpha}} \,.
\end{equation}
Excluding parameters (i.e. fixing these parameters) in Fisher analysis is convenient and we only need to remove the corresponding rows and columns without any further computation being required.
The extreme case is the \textit{conditional} uncertainty of a parameter given by $(F_{\alpha\alpha})^{-1/2}$.
Unless otherwise specified, the constraints on the parameters in this work are always the \textit{marginal} uncertainties.

Assuming the data vector $\datavh$ to be Gaussian, the likelihood function reads
\begin{equation}
\begin{split}
    \mathcal{L}(\datavh|\bm{\theta}) =\ &\frac{1}{(2\pi)^{{\rm dim}(\datavh)/2}\sqrt{|\covc|}} \\
    &\times \exp\left\{-\frac{1}{2}(\datavh-\datav)^\dagger \covc^{-1} (\datavh-\datav)\right\} \,,
\end{split}
\end{equation}
where ${\rm dim}(\datavh)$ is the length of $\datavh$, with the ensemble average $\datav\equiv\langle\datavh\rangle$ and covariance matrix
\begin{equation}
    \covc\equiv\langle(\datavh-\datav)(\datavh-\datav)^\dagger\rangle=\langle\datavh\datavh^\dagger\rangle - \datav\datav^\dagger \,,
\end{equation}
whose determinant is denoted as $|\covc|$.
Then the explicit expression of Fisher matrix given the Gaussian likelihood can be written as
\begin{equation} \label{eq:fisher_Gaussian}
    F_{\alpha\beta} = \frac{1}{2}{\rm Tr}[\covc^{-1}\covc_{,\alpha} \covc^{-1}\covc_{,\beta}] + \datav^\dagger_{,\alpha} \covc^{-1} \datav_{,\beta} \,.
\end{equation}

\subsection{Data vector} \label{subsec:data_vector}
In our Fisher analyses, we take the SH or SFB coefficients of the 2D or 3D fields as the data vector.
CMB lensing and galaxy overdensity fields have constant ensemble average (e.g. zero, depending on the definition) and hence the second term in Eq.~(\ref{eq:fisher_Gaussian}) vanishes.
We ignore the possible mode coupling of the angular multipoles $\ell$'s and simply use the fractional sky coverage $f_{\rm sky}$ to account for the loss of information due to partial sky survey footprint.
This approximation is reasonable given the large sky coverages with regular geometries that we consider.
In general, this coupling could be reduced by binning the modes or removed with the mode coupling matrix given the angular mask of the survey~\cite{Hivon2002}.

Then each $\ell$ contributes independently to the Fisher matrix, and the total information can be written as a summation
\begin{equation} \label{eq:fm}
    F_{\alpha\beta} = \sum_\ell \frac{(2\ell+1)f_{\rm sky}}{2} {\rm Tr}[\covc_\ell^{-1}\covc_{\ell,\alpha} \covc_\ell^{-1}\covc_{\ell,\beta}] \,,
\end{equation}
where the $2\ell+1$ factor results from the number of equivalent $m$ modes for each $\ell$, since the covariances of the coefficients are the power spectra, which do not depend on $m$ as shown in Section~\ref{subsec:power_spectra}.
This is the explicit form that is used in the Fisher analyses in this work.
Next let us look into $\datavh_{\ell m}$ and the corresponding $\covc_\ell$ for the two ways of decomposing 3D galaxy fields, TSH and SFB analyses.

First we consider the joint analysis of $\kappa$ and tomographic $g$ maps from one or multiple galaxy samples.
For each $\ell$ and $m$, the data vector reads
\begin{equation} \label{eq:D_lm_tomo}
    \datavh_{\ell m} = (\hat{\kappa}_{\ell m},\,\{\hat{g}^{ij}_{\ell m}\})^T \,,
\end{equation}
where the set $\{\hat{g}^{ij}_{\ell m}\}$ includes all the galaxy samples considered, indexed with $i$, and for each sample, $j$ denotes the redshift bins.
Notice that different galaxy samples or redshift bins could have different maximum $\ell$ given our discussion in Section~\ref{subsec:linear_modes} about the linear modes being included in the analysis, and hence the number of samples/bins included in $\datavh_{\ell m}$ could vary for different $\ell$.
It is slightly messy but still straightforward to understand since different $\ell$ modes contribute independently to the Fisher information, and we should have the freedom to decide what data to use for each $\ell$ as long as the choice is consistent for all the Fisher matrix elements.

The formalism is similar in SFB analysis except that for each galaxy sample we have multiple discrete radial modes instead of tomographic redshift bins.
The data vector can be written as
\begin{equation} \label{eq:D_lm_sfb}
    \datavh_{\ell m} = (\hat{\kappa}_{\ell m},\,\{\hat{\delta}^i_{\ell m n}\})^T \,,
\end{equation}
where $i$ denotes galaxy samples and $n$ is the index for discrete radial wavenumbers as discussed in Section~\ref{subsec:decomposition}.

For both TSH and SFB analyses, we consider the full covariance matrix of $\datavh_{\ell m}$.
The power spectra for any pair of SH or SFB coefficients in $\datavh_{\ell m}$ are computed using the expressions in Eqs.~(\ref{eq:cl_22}),~(\ref{eq:cl_23}) and~(\ref{eq:cl_33}).
For each $\ell$, the galaxy samples included and the number of radial modes for each sample could be different.

We use the Fisher matrix in Eq.~(\ref{eq:fm}) for the analyses in this work, while it is also helpful to implement an equivalent form as a double check, which is given as
\begin{equation} \label{eq:fm_ps}
    F_{\alpha\beta} = \sum_\ell \bm{d}_{\ell,\alpha}^\dagger \mathbf{M_\ell}^{-1} \bm{d}_{\ell,\beta} \,,
\end{equation}
where $\bm{d}_\ell$ is a vector consisting of all the power spectra, i.e. a stack of the upper triangular elements in $\covc_\ell$, and $\mathbf{M}_\ell$ is the Gaussian covariance matrix of $\bm{d}_\ell$, where a similar $(2\ell+1)f_{\rm sky}$ sampling factor is included as shown in Eq.~(\ref{eq:Gaussian_cov_ps}).
Eq.~(\ref{eq:fm_ps}) is sometimes referred as the Fisher matrix \textit{at power spectra level}, and it is mathematically equivalent as Eq.~(\ref{eq:fm}), see more discussion in~\cite{Bellomo2020,Hamimeche2008}.
It is worth being reminded that Eq.~(\ref{eq:fm_ps}) is not given by taking the power spectra vector (i.e. $\bm{d}_\ell$) as the Gaussian data vector in the general Gaussian Fisher matrix in Eq.~(\ref{eq:fisher_Gaussian}), where the first term would not vanish since $\mathbf{M}_\ell$ is also function of the parameters.
The reason that Eq.~(\ref{eq:fm_ps}) is not preferred for all the analyses in this work is that the size of $\mathbf{M_\ell}$ could be very large and the inversion would take much longer computational time than the inversion of $\covc_\ell$.
For example, for a certain $\ell$, consider the joint analysis of $\delta$ with $n$ radial modes and $\kappa$.
Then ${\rm dim}(\covc_\ell)=(n+1)\times (n+1)$ , ${\rm dim}(\bm{d}_\ell) = (n+1)(n+2)/2$, and ${\rm dim}(\mathbf{M}_\ell)={\rm dim}(\bm{d}_\ell)\times{\rm dim}(\bm{d}_\ell)$.
In our analyses, $n$ can be of order $\sim 100$.
Thus we only use Eq.~(\ref{eq:fm_ps}) as a double check and run it for a few cases.

\subsection{\label{subsec:prior_info}Prior information}
\begin{figure}
    \centering
    \includegraphics[width=\columnwidth]{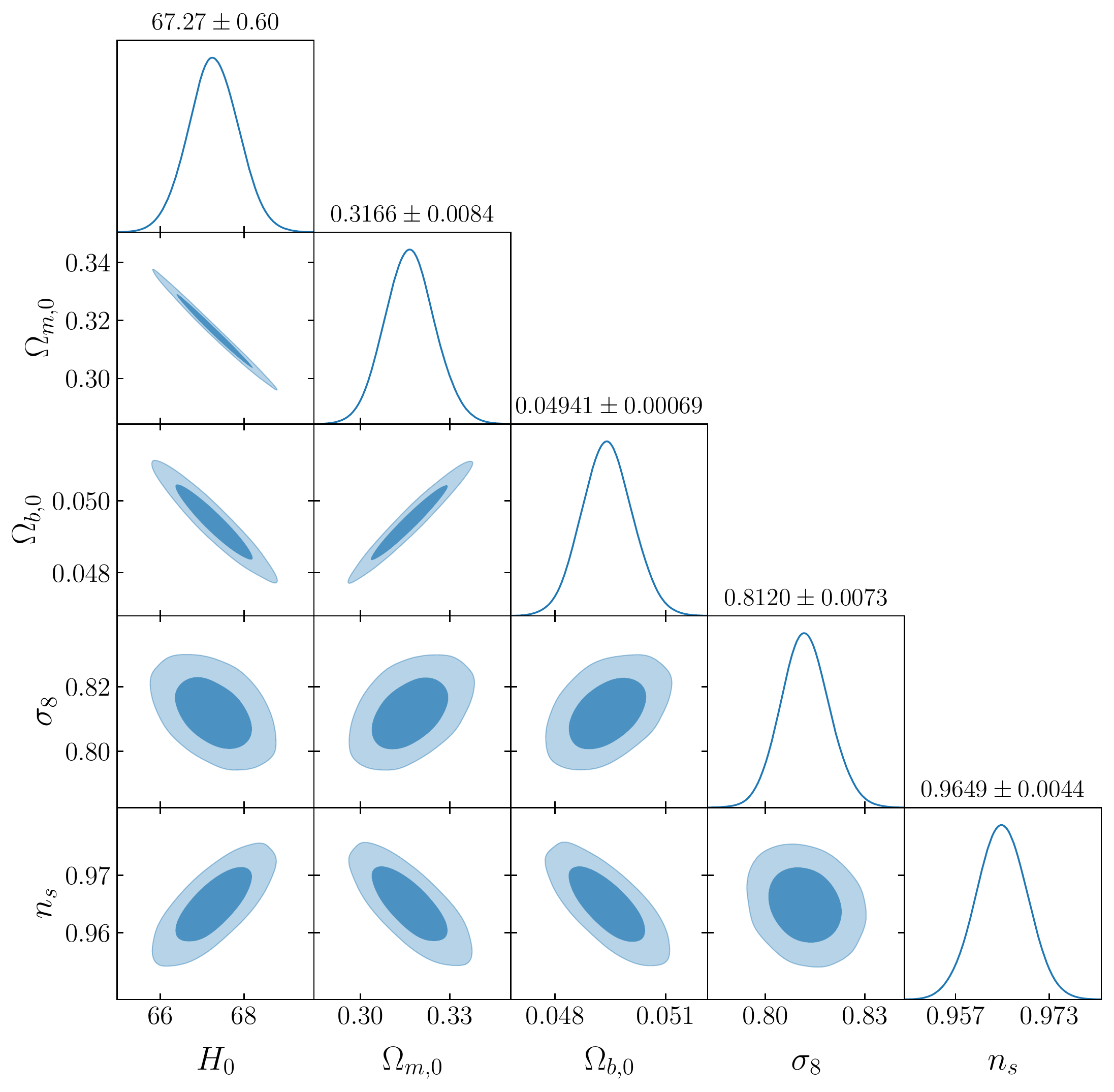}
    \caption{\label{fig:planck_prior}\textit{Planck} 2018 CMB TT,TE,EE+lowE constraints on the five background cosmological parameters considered in this work, recompiled from Monte Carlo chains provided using \textsc{GetDist}~\cite{Lewis2019,web-getdist}.
    Contours show $68\,\%$ and $95\,\%$ confidence regions. The corresponding covariance matrix will be used as prior information in our Fisher analyses, see Section~\ref{subsec:prior_info}.}
\end{figure}

For the five background cosmological parameters $\{H_0,\,\Omega_{m,0},\,\Omega_{b,0},\,\sigma_8,\,n_s\}$ considered, it would be helpful to include the prior information from \textit{Planck} 2018 CMB temperature and polarization data.
We use the \textit{Planck} TT,TE,EE+lowE constraints~\cite{Planck2018-cosmo}, and the covariance matrix for a subset of original and new derived parameters is reconstructed from the Monte Carlo chains provided at~\cite{web-planck-chains}.
We do not use the \textit{Planck} results including CMB lensing to avoid double counting information, since we have lensing in our Fisher analyses.
Using \textsc{GetDist}~\cite{Lewis2019,web-getdist}, the covariances for the five parameters are estimated, which are shown in Fig.~\ref{fig:planck_prior}.
As prior information, elements in the inverse of this covariance matrix are added to the corresponding Fisher matrix elements.

\section{\label{sec:results}Results and discussions}
In this section, we present and discuss the main results of our Fisher analyses.

\subsection{Power spectra}
\begin{table}
    \caption{\label{tab:clxx_snr}Total SNR of power spectra given by Eq.~(\ref{eq:snr_clxx}) for the galaxy samples decomposed in SFB or TSH basis.
    For TSH analyses of spectroscopic samples, a bin size $\Delta z=0.01$ is used.
    While for (spectro-)photometric samples, the bin sizes are limited by the redshift uncertainties (except for SPHEREx 1).
    These include the same linear modes that are used in the Fisher analyses, as discussed in Section~\ref{subsec:linear_modes}.}
    \begin{ruledtabular}
    \begin{tabular}{ldddd}
     & \multicolumn{4}{c}{SNR} \\
     \cmidrule(lr){2-5}
     & \multicolumn{1}{c}{$\bm{C}^{\kappa \delta}$} & \multicolumn{1}{c}{$\bm{C}^{\kappa g}$} & \multicolumn{1}{c}{$\bm{C}^{\delta \delta}$} & \multicolumn{1}{c}{$\bm{C}^{g g}$} \\
    \hline
    DESI BGS  & 22 & 21 & 263 & 251 \\
    DESI ELG  & 94 & 78 & 887 & 793 \\
    \textit{Euclid} & 140 & 116 & 1451 & 1330 \\
    LSST low-$z$ & 21 & 21 & 73 & 71 \\
    LSST high-$z$ & 38 & 37 & 178 & 156 \\
    SPHEREx 1 & 89 & 72 & 749 & 608 \\
    SPHEREx 2 & 78 & 78 & 583 & 513 \\
    SPHEREx 3 & 32 & 32 & 133 & 127 \\
    \end{tabular}
    \end{ruledtabular}
\end{table}

First as a simple check on our theoretical expressions and also numerical computations of the power spectra and noises for both SFB and TSH analyses, we estimate the total signal-to-noise ratio (SNR), which is given by
\begin{equation} \label{eq:snr_clxx}
    {\rm SNR}\left(\bm{C}^{x y}\right) = \left[ \sum_{\ell} (\bm{C}_\ell^{x y})^T {\rm Cov}_\ell^{-1} \bm{C}_\ell^{x y} \right]^{1/2} \,,
\end{equation}
where we sum over all the $\ell$ modes included in Fisher analyses, $\bm{C}_\ell^{x y}$ is the signal (i.e. without noise) vector and ${\rm Cov}_\ell$ is the Gaussian covariance matrix of $\bm{C}_\ell^{x y}$, with a general expression given by Eq.~(\ref{eq:Gaussian_cov_ps}).
The pair of fields $x y$ can be $\kappa \delta$ or $\kappa g$ for the CMB lensing and galaxy (in SFB or TSH basis) cross-power spectra, where for each $\ell$, $\bm{C}_\ell^{x y}$ is a 1D vector consists of power spectra of all the radial modes or redshift bins.
The fields $x y$ can also be $\delta\delta$ or $g g$ for the galaxy auto-power spectra, which are matrices for each $\ell$ and the vector $\bm{C}_\ell^{x y}$ is a stack of the upper triangular elements.
Notice that Eq.~(\ref{eq:snr_clxx}) is in a similar form as Eq.~(\ref{eq:fm_ps}), which is equivalent as Eq.~(\ref{eq:fm}).
Therefore for $\delta\delta$ and $g g$ where ${\rm Cov}_\ell$ can be too large, instead of Eq.~(\ref{eq:snr_clxx}), we use its equivalent expression as Eq.~(\ref{eq:fm}) to speed up the computation.
The results are summarized in Table~\ref{tab:clxx_snr}.
We can see that with their own linear modes, SFB give higher SNRs than TSH for both the auto-power spectra of galaxies and the cross-power spectra with CMB lensing.
\begin{figure}
    \centering
    \includegraphics[width=\columnwidth]{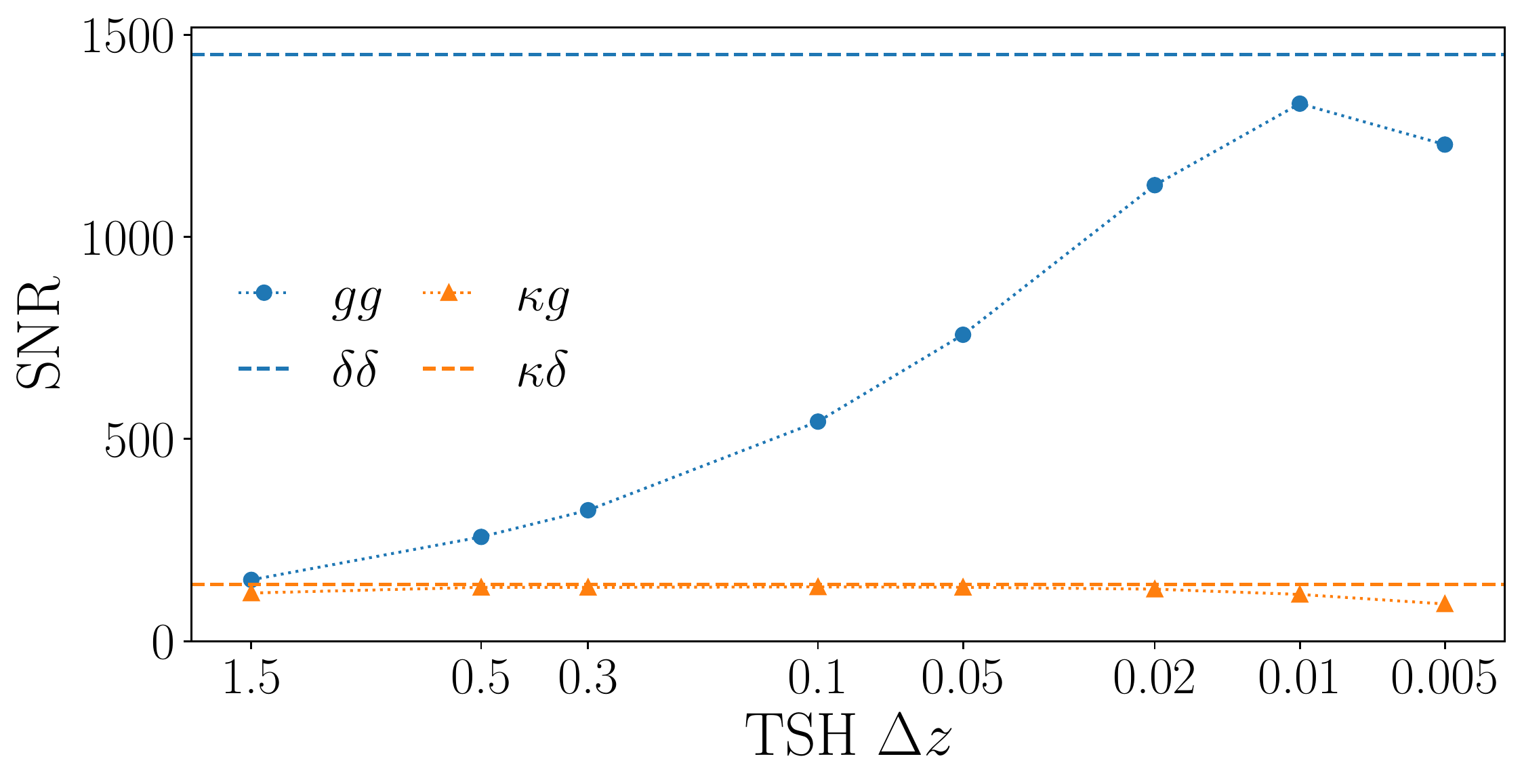}
    \caption{\label{fig:snr_tsh_bins_euclid_li}SNRs of TSH power spectra $\bm{C}^{gg}$ and $\bm{C}^{\kappa g}$ with different bin sizes for the \textit{Euclid} galaxy sample.
    SNRs of SFB power spectra $\bm{C}^{\delta\delta}$ and $\bm{C}^{\kappa \delta}$ are also shown as horizontal dashed lines for reference.}
\end{figure}

Of course the SNRs of TSH power spectra could depend on the number of tomographic redshift bins.
We take the \textit{Euclid} galaxy sample as an example and try different bin sizes, with the resulted SNRs shown in Fig.~\ref{fig:snr_tsh_bins_euclid_li}.
It is interesting that for the galaxy auto-power spectrum, instead of continuously increasing, SNRs of TSH with different bin sizes peak around $\Delta z = 0.01$, which is still lower than the SNR of SFB.
This is mainly due to the linear requirement which excludes more nonlinear modes when very small TSH bin sizes are used, as shown in Fig.~\ref{fig:tsh_linear_lmax_euclid}.
While for the cross-power spectrum with CMB lensing, TSH has similar SNRs as SFB for most bin sizes, which then decreases for very small bin sizes.
The SNR only tells us the overall strength of the power spectra signal, and a higher SNR does not guarantee better constraints on certain parameters, which could change the power spectra in different ways instead of simply tuning the amplitude.
In the following sections, we will discuss the constraints on different parameters, and how those given by TSH analyses depend on the bin size.

\subsection{\label{subsec:constraints_on_parameters}Constraints on parameters}
\begin{table*}
    \caption{\label{tab:par_constraints}Fisher forecasts of marginal constraints on the PNG parameter $f_{\rm NL}$ and the RSD exponent $\gamma$ for three progressive scenarios: galaxy only, joint analyses with CMB lensing, and further adding \textit{Planck} CMB temperature and polarization prior information.
    Results for galaxy samples analyzed in both SFB and TSH bases are included.
    For spectroscopic samples, a TSH bin size $\Delta z=0.01$ is used.
    For (spectro-)photometric samples (except SPHEREx 1), the constraints on $\gamma$ are very poor due to redshift uncertainties and thus not included.
    See Section~\ref{subsec:constraints_on_parameters} for more discussions.
    Note that $\ell_{\rm min}=2$ is used for these analyses, we discuss the dependence on $\ell_{\rm min}$ in Section~\ref{subsec:ell_min_dependence}.
    }
    \begin{ruledtabular}
    \begin{tabular}{l dddd | dddd | dddd}
         & \multicolumn{4}{c}{Galaxy only} & \multicolumn{4}{c}{$\times$ CMB lensing} & \multicolumn{4}{c}{+ \textit{Planck} prior} \\
        \cmidrule(lr){2-5}\cmidrule(lr){6-9}\cmidrule(lr){10-13}
         & \multicolumn{2}{c}{$\sigma(f_{\rm NL})$} & \multicolumn{2}{c}{$\sigma(\gamma)$} & \multicolumn{2}{c}{$\sigma(f_{\rm NL})$} & \multicolumn{2}{c}{$\sigma(\gamma)$} & \multicolumn{2}{c}{$\sigma(f_{\rm NL})$} & \multicolumn{2}{c}{$\sigma(\gamma)$} \\
        \cmidrule(lr){2-3}\cmidrule(lr){4-5}\cmidrule(lr){6-7}\cmidrule(lr){8-9}\cmidrule(lr){10-11}\cmidrule(lr){12-13}
         & \multicolumn{1}{c}{SFB} & \multicolumn{1}{c}{TSH} & \multicolumn{1}{c}{SFB} & \multicolumn{1}{c}{TSH} & \multicolumn{1}{c}{SFB} & \multicolumn{1}{c}{TSH} & \multicolumn{1}{c}{SFB} & \multicolumn{1}{c}{TSH} & \multicolumn{1}{c}{SFB} & \multicolumn{1}{c}{TSH} & \multicolumn{1}{c}{SFB} & \multicolumn{1}{c}{TSH} \\
        \cmidrule(lr){2-5}\cmidrule(lr){6-9}\cmidrule(lr){10-13}
        DESI BGS      & 45.1 & 44.9 & 0.19 & 0.22 & 40.1 & 40.4 & 0.038 & 0.043 & 33.9 & 34.2 & 0.029 & 0.033 \\
        DESI ELG      & 7.9 & 8.2 & 0.067 & 0.078 & 7.6 & 7.8 & 0.021 & 0.026 & 7.2 & 7.4 & 0.017 & 0.020  \\
        DESI BGS+ELG  & 7.8 & 8.0 & 0.038 & 0.049 & 7.0 & 7.2 & 0.019 & 0.024 & 6.7 & 6.9 & 0.015 & 0.018  \\
        \addlinespace[4.8pt]
        \textit{Euclid} & 4.6 & 4.7 & 0.034 & 0.040 & 4.4 & 4.5 & 0.015 & 0.019 & 4.2 & 4.3 & 0.012 & 0.014  \\
        \addlinespace[4.8pt]
        LSST low-$z$ & 6.2 & 6.5 & $-$ & $-$ & 3.3 & 3.4 & $-$ & $-$ & 2.6 & 2.6 & $-$ & $-$ \\
        LSST high-$z$ & 1.2 & 1.3 & $-$ & $-$ & 0.9 & 1.0 & $-$ & $-$ & 0.6 & 0.6 & $-$ & $-$ \\
        LSST all & 0.8 & 0.9 & $-$ & $-$ & 0.7 & 0.7 & $-$ & $-$ & 0.5 & 0.5 & $-$ & $-$ \\
        \addlinespace[4.8pt]
        SPHEREx 1 & 4.8 & 5.0 & 0.043 & 0.055 & 3.9 & 4 & 0.019 & 0.028 & 3.8 & 3.8 & 0.018 & 0.026 \\
        SPHEREx 2 & 2.9 & 3.0 & $-$ & $-$ & 2.5 & 2.6 & $-$ & $-$ & 2.3 & 2.4 & $-$ & $-$ \\
        SPHEREx 3 & 5.4 & 5.8 & $-$ & $-$ & 3.0 & 3.2 & $-$ & $-$ & 2.5 & 2.6 & $-$ & $-$ \\
        SPHEREx 1-3 & 2.2 & 2.3 & $-$ & $-$ & 1.9 & 2.0 & $-$ & $-$ & 1.9 & 1.9 & $-$ & $-$ \\
    \end{tabular}
    \end{ruledtabular}
\end{table*}
We summarize the Fisher constraints on the two parameters of primary interest, $f_{\rm NL}$ and $\gamma$, in Table~\ref{tab:par_constraints} for three scenarios: galaxy only, joint analyses with CMB lensing, and further adding prior information from \textit{Planck} CMB temperature and polarization.
For the (spectro-)photometric samples with high redshift uncertainties, we do not report the poor constraints on $\gamma$, which are not comparable to the constraints given by spectroscopic samples.
The only exception is the SPHEREx 1 sample, whose redshift uncertainty is actually low enough to be treated as a spectroscopic sample.
As mentioned in Section~\ref{subsec:parameters}, these are the constraints with other parameters being marginalized, including five background cosmological parameters $\{H_0,\,\Omega_{m,0},\,\Omega_{b,0},\,\sigma_8,\,n_s\}$ and two nuisance parameters $\{A_b,\,s\}$.
\begin{figure*}
    \centering
    \includegraphics[width=\textwidth]{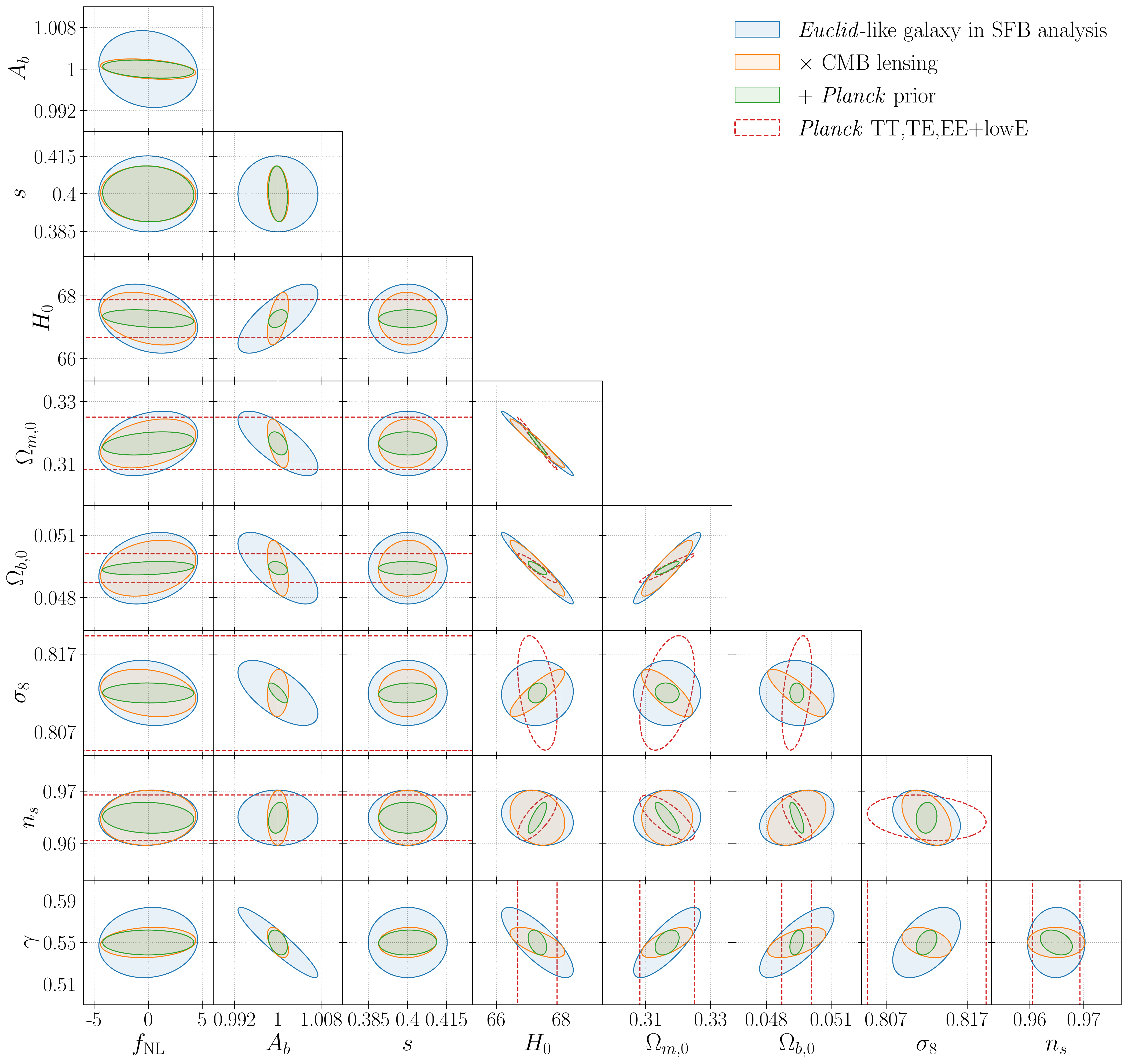}
    \caption{\label{fig:euclid_constraints_all_li}Fisher constraints on the PNG parameter $f_{\rm NL}$, the RSD exponent $\gamma$, five background cosmological parameters $\{H_0,\,\Omega_{m,0},\,\Omega_{b,0},\,\sigma_8,\,n_s\}$, and two nuisance parameters $\{A_b,\,s\}$ with the \textit{Euclid}-like galaxy sample.
    The contours shown are $1\,\sigma$ ($68\,\%$) confidence regions.
    We consider the constraints with galaxy only in SFB basis (blue), joint analysis with CMB lensing (orange), and further adding the \textit{Planck} CMB TT,TE,EE+lowE prior (green).
    Note that the CMB prior (dashed red lines) only contains information for the five background cosmological parameters.}
\end{figure*}
Besides these marginal constraints on $f_{\rm NL}$ and $\gamma$, to further look into the correlations between any pair of free parameters in the Fisher analysis, as an example, we show the full constraints with the \textit{Euclid} galaxy sample in Fig.~\ref{fig:euclid_constraints_all_li}.

In Table~\ref{tab:par_constraints}, besides the results for the individual galaxy samples listed back in Table~\ref{tab:galaxy_surveys}, we also show the results given by combining samples from the same survey (e.g. DESI BGS and ELG) in the data vector in Eq.~(\ref{eq:D_lm_tomo}) for TSH and Eq.~(\ref{eq:D_lm_sfb}) for SFB.
The covariances between samples are fully included since samples from the same survey are observing the same angular patch of the sky.
While for samples with different sky coverages, Fisher information for overlapping and non-overlapping regions should be calculated separately with and without covariances, and then combined.
However, the footprint overlap between different surveys depends closely on the observation details of these future surveys, which are not very clear at this stage.
Therefore here we do not discuss the combination of galaxy samples from different surveys.

For $f_{\rm NL}$, we notice that TSH analyses give similar constraints as SFB.
This means that the bin size should be small enough in recovering large radial scales where PNG is more significant, and we discuss more about this in Section~\ref{subsec:TSH_bin_size}.
Recall that one of our motivations is to check how radial information contributes to constraining $f_{\rm NL}$.
The extreme cases are SFB analysis where radial information is fully considered, and TSH analysis with only one bin where most if not all radial information is lost in the projection.
For all the galaxy samples considered, compared to the TSH analysis with only one bin, SFB could be better by a factor of 3 to 12.
We also tried TSH with two bins, and $\sigma(f_{\rm NL})$ gets much tighter compared to the one bin case, while SFB could still be better by a factor of 2 to 3.
These improvement factors vary for different surveys, while the general conclusion is that large radial scales does contribute significantly to constraining $f_{\rm NL}$.
Joint analysis with CMB lensing improves $\sigma(f_{\rm NL})$ more when the cross-correlation (Fig.~\ref{fig:clkg_corr_coef}) is stronger at low $\ell$s, e.g. for DESI BGS.
Besides these marginal constraints, in Fig.~\ref{fig:euclid_constraints_all_li}, we can see that the covariances between $f_{\rm NL}$ and other parameters are not strong.
This is one of the reasons that we do not see the improvements with CMB lensing that are as significant as those shown in~\cite{Schmittfull2018}.
Instead of considering background cosmological parameters, they introduced a fake $f_{\rm NL}$ parameter to the matter power spectrum that mimics the real $f_{\rm NL}$ in scale-dependence.
This resulted in a degeneracy that is much stronger than it should be, and therefore CMB lensing became more important in reducing that.

For constraining $\gamma$, with CMB lensing included, we find significant improvements by a factor of 2 to 5 for different samples depending on their redshift ranges and scales included in the Fisher analyses.
These improvements on $\sigma(\gamma)$ mainly come from the mitigation of degeneracies with other parameters, which can be seen from the shapes and orientations of the confidence regions shown in Fig.~\ref{fig:euclid_constraints_all_li}.
With galaxy only, $\gamma$ is strongly correlated with the clustering bias $A_b$, which is the well-known RSD-bias degeneracy since it is roughly the sum of $f\sigma_8$ and $b_g\sigma_8$ that determines the overall amplitude of the power spectrum.
Both $\gamma$ and $A_b$ are also correlated with some of the background cosmological parameters.
After CMB lensing is included, these covariances are reduced, especially between $\gamma$ and $A_b$.
On the other hand, for the comparison between SFB and TSH methods, we get better constraints on $\gamma$ with SFB.
This indicates that even with a small enough bin size, linear TSH modes still contain less radial information than SFB.
More discussions are included in Section~\ref{subsec:TSH_bin_size} below.

For the background cosmological parameters, the constraints are also improved with the joint analysis with CMB lensing and also the addition of CMB temperature and polarization prior information.
Another interesting point to notice is that the galaxy magnification bias is almost not degenerate with any other parameters.
As a result, for the samples at lower redshifts, we do not observe much difference with $s$ being fixed or marginalized, even with different fiducial $s$ values we tried in the range $0.1-0.7$.
The only exception is the LSST high-$z$ sample which covers redshift $2<z<5$, for which we do observe relative differences of dozens of percent in $\sigma(f_{\rm NL})$ with different fiducial $s$ values being used.
This is understandable considering that magnification bias is caused by the foreground lensing, to which galaxy samples at higher redshifts might be more sensitive.
While with CMB lensing included, $\sigma(f_{\rm NL})$ becomes much less dependent on $s$, which is another advantage of the joint analysis.

\subsection{Dependence on the minimum angular multipole} \label{subsec:ell_min_dependence}
In our main analyses, we use the minimum angular multipole $\ell_{\rm min}=2$ for all the surveys, which is reasonable given the large sky coverage ($f_{\rm sky}$ in Table~\ref{tab:galaxy_surveys}) of these surveys.
However, even though spatially accessible, these very large scales have always been challenged by systematics, which makes them excluded from practical analyses.
Therefore in this part we discuss the dependence of the parameter constraints on the $\ell_{\rm min}$ used in Fisher analyses.

\begin{figure}
    \centering
    \includegraphics[width=\columnwidth]{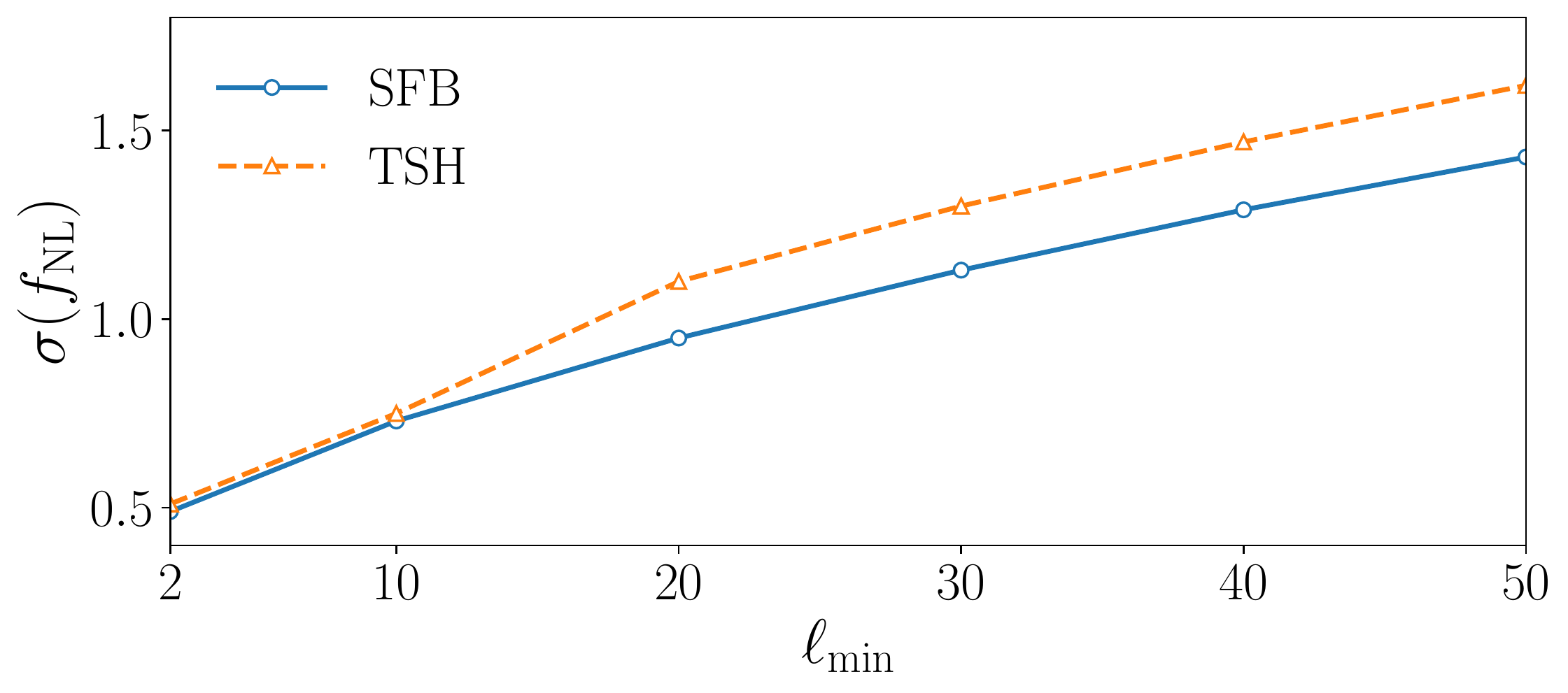}
    \caption{\label{fig:lmin_lsst}Dependence of $\sigma(f_{\rm NL})$ on the minimum multipole $\ell_{\rm min}$ used for Fisher analyses, shown for the LSST sample (joint analyses with CMB lensing and with \textit{Planck} prior added) as an example.}
\end{figure}

In Fig.~\ref{fig:lmin_lsst}, taking the LSST sample (which gives the best constraint on $f_{\rm NL}$) as an example, we show the dependence of $\sigma(f_{\rm NL})$ on $\ell_{\rm min}$ used in the Fisher analyses.
We can see that $f_{\rm NL}$ is very sensitive to low $\ell$ modes (i.e. large scales) given its $k^{-2}$ scale dependence.
Using $\ell_{\rm min}=50$ could increase the uncertainty in $f_{\rm NL}$ by a few factors (e.g. 3 for this LSST example) compared to using $\ell_{\rm min}=2$.
Thus for future surveys dedicated to constraining $f_{\rm NL}$, it would be very helpful to identify and reduce large scale systematics.
While for the growth rate exponent $\gamma$, the constraint is less sensitive to $\ell_{\rm min}$.
For spectroscopic surveys like the \textit{Euclid} sample, using $\ell_{\rm min}=100$ only increases $\sigma(\gamma)$ by around $20\,\%$.

\subsection{\label{subsec:TSH_bin_size}Dependence of TSH constraints on the bin size}
As discussed in Section~\ref{subsec:linear_modes}, the modes included in Fisher analyses are determined based on the linear requirement of the SFB and TSH power spectra.
For TSH analysis, the maximum linear angular multipoles also depend on the bin size, as shown in Fig.~\ref{fig:tsh_linear_lmax_euclid} for the \textit{Euclid} sample as an example.
In the discussions above, we use the bin size $\Delta z = 0.01$ in TSH analyses, and here we discuss how this optimal value is found.
\begin{figure}
    \centering
    \includegraphics[width=\columnwidth]{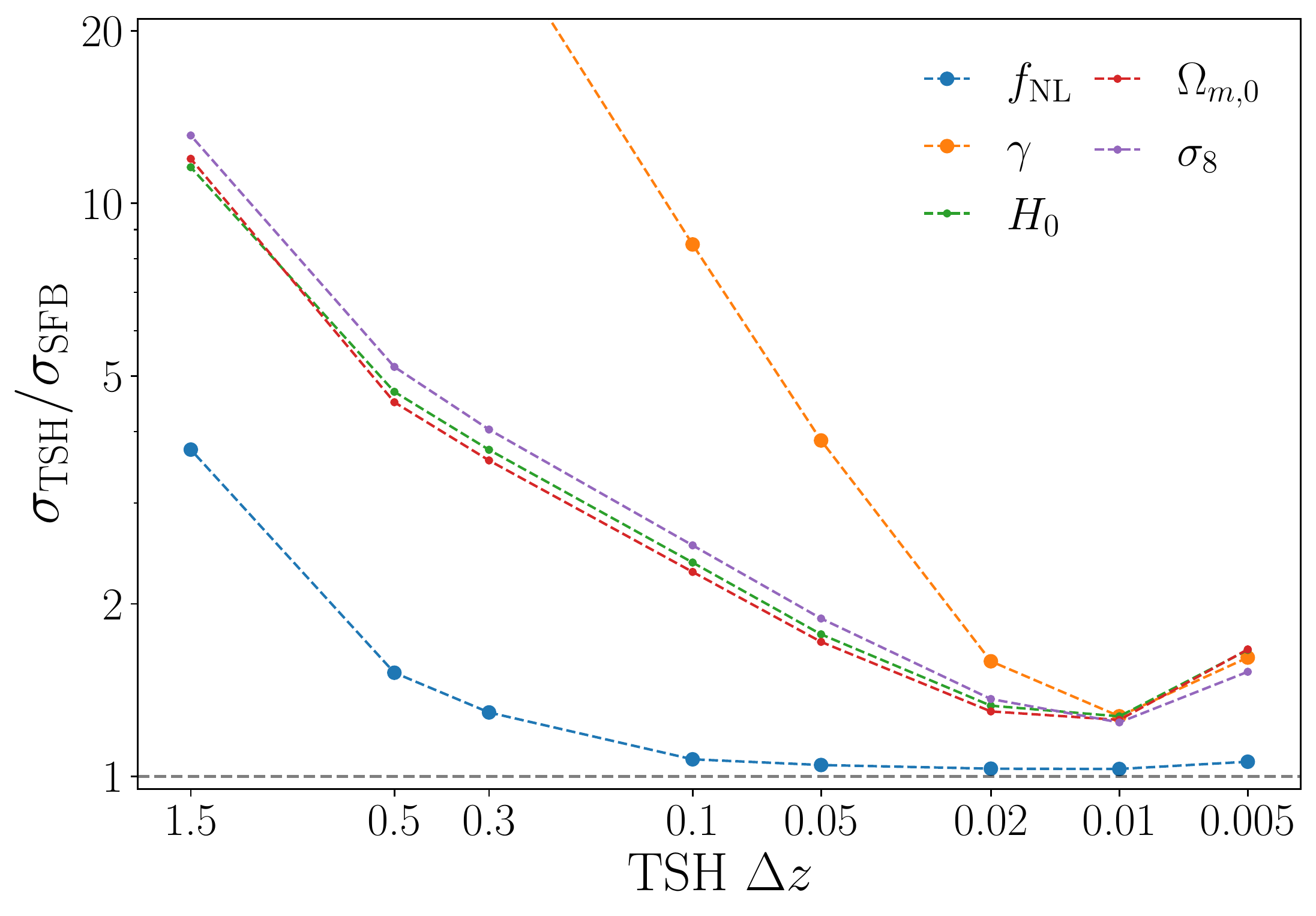}
    \caption{\label{fig:params_tsh_bins_euclid_li}The ratios between parameter constraints given by TSH and SFB analyses of the \textit{Euclid}-like galaxy sample, with CMB lensing included.
    To get a more straightforward idea about the dependence of background cosmological parameters on the bin size $\Delta z$ in TSH analysis, \textit{Planck} CMB prior is not added.
    See Section~\ref{subsec:TSH_bin_size} for more details.}
\end{figure}
Fig.~\ref{fig:params_tsh_bins_euclid_li} shows the TSH constraints on parameters with different bin sizes, where the values are shown as ratios to the SFB constraints.
Similar as the SNR of the galaxy power spectrum that peaks around $\Delta z = 0.01$ as shown in Fig.~\ref{fig:snr_tsh_bins_euclid_li}, the tightest constraints on the parameters given by TSH are also achieved around $\Delta z=0.01$.
Given that RSD is a purely radial effect, $\sigma(\gamma)$ is more sensitive to the bin size than other parameters.
On the other hand, compared with other parameters, $\sigma(f_{\rm NL})$ requires fewer bins to reach the SFB constraint since it is more sensitive to large scales and the additional information from very small bins does not contribute a lot.
Similar discussion in a simplified cubic box geometry can be found in~\cite{Ferraro2015}, where it is shown that analyzing a 3D survey as a 2D map will lose a factor greater than 2 in SNR, consistent with our results.

Fig.~\ref{fig:params_tsh_bins_euclid_li} and~\ref{fig:snr_tsh_bins_euclid_li} are shown for the \textit{Euclid} sample, but the bin size dependence and the optimal bin size are similar for DESI BGS, ELG or SPHEREx 1 galaxy samples we consider.
For other (spectro-)photometric samples, the optimal bin sizes could not be achieved since they are smaller than the redshift uncertainties.
The optimal bin size depends on many details of the survey including the number density which determines the shot noise level, and also the redshift distribution.
We tried finer sampling of the bin size, and the optimal values are not exactly the same for different surveys.
For example, for DESI BGS, the optimal $\Delta z$ is closer to $0.008$, which is smaller than $0.01$.
While for DESI ELG and \textit{Euclid}, the finer optimal values are slightly lower than $0.01$.
We also tuned the number density with other configurations being fixed and noticed that, the advantage of SFB is stronger when the shot noise is lower.

\subsection{\label{subsec:cov_modes_bins}Covariances between radial modes in SFB or redshift bins in TSH}
In this part, we consider the importance of the covariances between radial modes in SFB or redshift bins in TSH.
In principle, all the covariances should be included as part of the total information, while sometimes people might ignore them for simplicity.
Therefore, it is worth being discussed how the constraints on parameters of interest would change with or without the covariances.

In TSH basis, the cross-correlations between tomographic redshift bins contain wealthy information from the radial direction.
Therefore excluding these covariances in the Fisher analysis could result in worse constraints on the parameters.
The significance of off-diagonal elements in a covariance matrix $\mathbf{C}$ can be quantified with the correlation matrix, whose elements are given by
\begin{equation} \label{eq:corr_def}
    {\rm Corr}(\mathbf{C})_{ij} = \frac{C_{ij}}{\sqrt{C_{ii}C_{jj}}} \,,
\end{equation}
where in this case ${C}_{ij}$ corresponds to Eq.~(\ref{eq:cl_22}) and (\ref{eq:cl_33}) for TSH and SFB galaxy power spectra respectively. 
\begin{figure}
    \centering
    \includegraphics[width=\columnwidth]{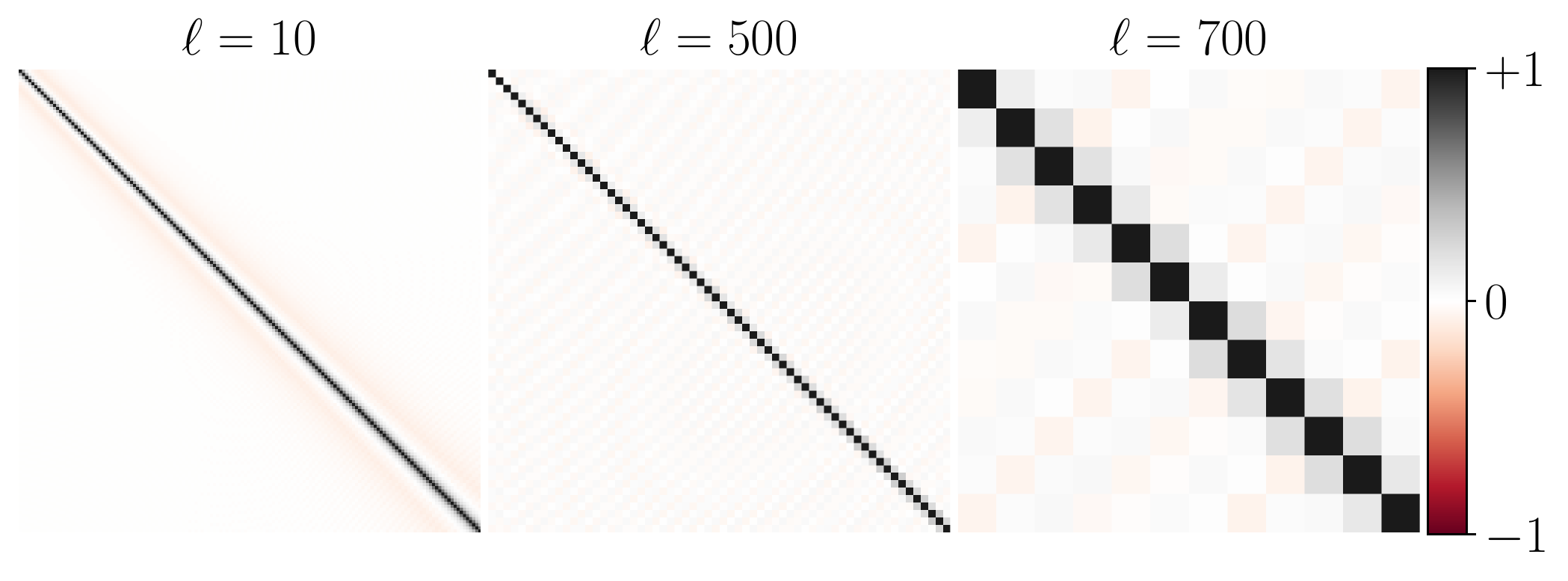}
    \caption{\label{fig:clgg_corr_euclid_li}Correlation matrices of TSH galaxy power spectra for $\ell=10,\,500,\,700$, assuming the \textit{Euclid} sample.
    The upper left corner corresponds to lower redshift bins.}
\end{figure}
As an example, we show the correlation matrices of TSH power spectra of the \textit{Euclid} galaxy sample for a few $\ell$'s in Fig.~\ref{fig:clgg_corr_euclid_li}.
We can see there are non-negligible correlations between redshift bins for both high and low $\ell$s, given the small bin size $\Delta z=0.01$ we use.
For the galaxy samples we consider, with the covariances removed, we noticed that both $\sigma(f_{\rm NL})$ and $\sigma(\gamma)$ could be larger approximately by a factor of 1.5 to 2.5.
Therefore as expected, it is crucial to consider the covariances in tomographic analysis in order to make use of the valuable radial information.

\begin{figure}
    \centering
    \includegraphics[width=\columnwidth]{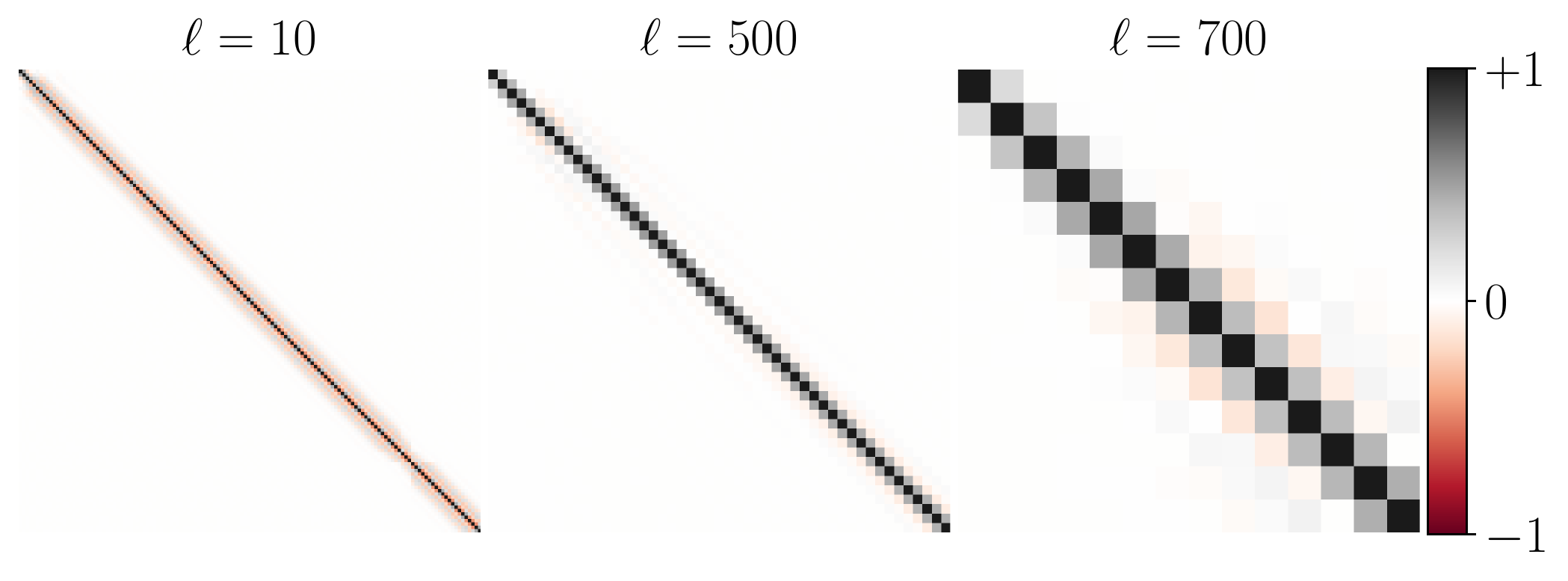}
    \caption{\label{fig:cldd_corr_euclid_li}Similar as Fig.~\ref{fig:clgg_corr_euclid_li} but for SFB galaxy power spectra. The upper left corner corresponds to lower $k_{\ell n}$ modes.}
\end{figure}
While in SFB analysis, it is not straightforward to analytically predict how the covariances between discrete radial modes would change the constraints.
We show a few correlation matrices in Fig.~\ref{fig:cldd_corr_euclid_li}.
Those non-zero off-diagonal covariances are mainly caused by the redshift dependence of the galaxy field and also the boundary condition, where the orthogonality relations of the radial basis functions no longer hold.
For the galaxy samples we consider, $\sigma(f_{\rm NL})$ could be larger by $10\,\%$ to $40\,\%$ due to these covariances.
While for $\gamma$, which is more sensitive to the radial information, the constraints could be either better or worse by dozens of percent depending on the specific sample.
Thus it is important to consider the covariances between radial modes in SFB power spectra analysis, given the significant impact on the parameter constraints.

\section{\label{sec:conclusions}Conclusions}
As observables tracing the same matter field, cross-correlating CMB lensing and galaxy clustering is powerful in reducing the sample variance on large scales and also mitigating the degeneracies between galaxy-only and standard $\Lambda$CDM cosmological parameters.
Compared with Cartesian $P(\bm{k})$ analysis, decomposing 3D spherical galaxy field in SFB basis is a more natural choice for large scales.
This also makes it straightforward to be cross-correlated with 2D CMB lensing map in SH basis.
Motivated by this SFB analysis that maintains the radial information, we investigate the constraints on the PNG parameter $f_{\rm NL}$ and the RSD exponent $\gamma$ by performing Fisher forecasts for galaxy setups that mimic a few future surveys.
In these Fisher analyses, we also marginalize over five $\Lambda$CDM cosmological parameters and two nuisance parameters that accounts for clustering bias and magnification bias.

We consider the linear modes that are defined based on their own power spectra in SFB and TSH analyses.
We avoid doing these by converting from the 3D linear scale as has been done in some previous work, since 3D wavenumbers are actually mixed in SFB and TSH analyses, and it is hard to perform the conversion accurately.
For TSH analysis, a direct result is that for a much smaller bin size, maximum $\ell$ for each bin would be lower since those modes that become more nonlinear due to the small bin size are excluded.
In general, we find that SFB works better than TSH in maintaining the linear modes and therefore gives more information in constraining parameters.

For $f_{\rm NL}$, thanks to the contribution from large radial scales, SFB gives tighter constraints by a factor of 3 to 12 compared to TSH analysis with only one bin, where radial information is mostly lost.
Since PNG is only significant on large scales, decreasing the bin size in TSH analysis could improve $\sigma(f_{\rm NL})$ but would not give better results than SFB analysis where large radial scales are clearly included.
We also notice that in SFB analysis or TSH analysis with a large number of bins, CMB lensing does improve $\sigma(f_{\rm NL})$ but not significantly since $f_{\rm NL}$ is only weakly degenerate with other cosmological parameters, and also radial scales contribute more modes to reducing the sample variance than CMB lensing.
For the galaxy samples considered, compared with analyzing galaxy only, joint analysis with CMB lensing could improve $\sigma(\gamma)$ by a factor of 2 to 5.
This is mainly contributed by reducing the degeneracies between $\gamma$ and other parameters, especially the clustering bias.

For the magnification bias $s$ due to foreground lensing, its degeneracies with other parameters are found to be very weak while different fiducial values could change $\sigma(f_{\rm NL})$ by dozens of percent for analyzing high redshift galaxy samples.
However, using the joint analysis with CMB lensing, $\sigma(f_{\rm NL})$ becomes more robust and these changes due to fiducial $s$ values reduce to only a few percent.
Therefore, for analyzing high redshift galaxy datasets, it might be necessary to consider free $s$ parameter in a proper prior range if CMB lensing is not included.

Both SFB and TSH methods have their own advantages and limitations.
In SFB basis, it is more convenient to decompose a 3D field without losing information, especially for large radial scales.
However, the sacrifice is that the information from different redshifts is mixed in the radial integral, which is an inevitable result of observing the light cone.
In TSH basis, it is easier to study the redshift evolution of the field, but the radial information may not be well reconstructed even with a large amount of modes.
Therefore, which method to use depends on the parameters of interest.
For example, if the primary goal is to constrain $f_{\rm NL}$, then TSH analysis with a moderate number of bins should suffice.
Besides, it is worth mentioning that for constraining $f_{\rm NL}$ in the 3D $P(k)$ analysis, an optimal redshift weighting method has been shown to be helpful in reducing the uncertainty (see e.g.~\cite{Mueller2019,Castorina2019,Mueller2021} for the application on the eBOSS data).
This might also be an interesting aspect to consider when the SFB formalism is used to analyze observed data in the future.

In general, our Fisher forecasts show that joint analyses of future CMB lensing and galaxy surveys in SFB basis are very promising in constraining PNG and RSD, which are probes of inflation and gravity models respectively.
For future large spectroscopic surveys like DESI BGS+ELG or \textit{Euclid}, we would be able to constrain $\gamma$ to $\sim 3\,\%$ precision using their linear scales.
For high redshift photometric samples like LSST, $\sigma(f_{\rm NL}) < 1$ can be achieved as long as $\ell_{\rm min}\simeq 20$ are free of possible large-scale systematics.
However, to use either SFB or TSH formalism for data analyses of future surveys, besides the estimator that has been discussed in~\cite{Leistedt2012,Gebhardt2021}, it is still necessary to improve the numerical algorithm of computing the theoretical power spectra since they would have to be evaluated at each MCMC step.
Besides the \textsc{FFTLog} algorithm mentioned in Appendix~\ref{sec:num_cal}, another promising solution is to extend the emulators (see e.g.~\cite{Arico2021,Mancini2021}) to the SFB power spectra.

\begin{acknowledgments}
We thank Simone Ferraro and Colin Hill for their helpful feedback on an early version of the manuscript.
We also thank Yue Shi for helpful discussions.
YZ was supported by a James Arthur Graduate Award from NYU GSAS.
ARP was supported by NASA under award numbers
80NSSC18K1014 and NNH17ZDA001N, as well as the Simons Foundation.
The numerical computations in this work were performed in part using the NYU Greene High Performance Computing cluster.

\end{acknowledgments}

\appendix

\section{\label{sec:orthogonality}Orthogonality relations}
In this Appendix, we present a brief review of the orthogonality relations satisfied by the radial basis functions in SFB decomposition.
Following the discussion in~\cite{Fisher1995}, we derive the normalization factors under different boundary conditions (BCs).
The spherical Bessel functions are defined through the differential equation
\begin{equation} \label{eq:diff_eqn}
    \frac{1}{r} \frac{d^2}{dr^2} \left[ r f_\ell(kr) \right] = \left[ \frac{\ell(\ell+1)}{r^2} - k^2 \right] f_\ell(kr)\,,
\end{equation}
where $f_\ell(kr)$ can be any linear combination of $j_\ell(kr)$ and $y_\ell(kr)$, the spherical Bessel functions of first and second kind.
Applying the operation
\begin{equation}
    \int_{r_1}^{r_2} dr\, r^2 f_\ell(k'r)
\end{equation}
to both sides of Eq.~(\ref{eq:diff_eqn}) and removing the symmetric terms in $k$ and $k'$ by doing the subtraction with $k$ and $k'$ interchanged, we are left with
\begin{equation} \label{eq:inte_eqn}
\begin{split}
    \int_{r_1}^{r_2} &dr\, r^2 f_\ell(kr) f_\ell(k'r) \\
    &= \frac{\left.r^2\left[ k' f_\ell(kr) f_\ell'(k'r) - k f_\ell(k'r) f_\ell'(kr) \right]\right|_{r_1}^{r_2}}{k^2-k'^2} \,,
\end{split}
\end{equation}
where $r_1$ and $r_2$ are the lower and upper radial boundaries.
We can see that for discrete $k=k_{\ell n}$ and $k'=k_{\ell n'}$ values determined with either \textit{Dirichlet}
\begin{equation} \label{eq:Dirichlet_BC}
    f_\ell(k_{\ell n}r_{1,2}) = 0
\end{equation}
or \textit{Neumann}
\begin{equation} \label{eq:Neumann_BC}
     f_\ell'(k_{\ell n}r_{1,2}) = 0
\end{equation}
BC, the numerator of Eq.~(\ref{eq:inte_eqn}) evaluated at the boundaries are zero.
Then Eq.~(\ref{eq:inte_eqn}) can be written as a orthogonality relation
\begin{equation} \label{eq:radial_ortho}
    \int_{r_1}^{r_2} dr\, r^2 f_\ell(k_{\ell n}r) f_\ell(k_{\ell n'}r) = \tau_{\ell n}\delta_{nn'}^{\rm K} \,,
\end{equation}
where the normalization factor $\tau_{\ell n}$ for $n=n'$ can be determined by taking the limit $k \rightarrow k'$ on the RHS of Eq.~(\ref{eq:inte_eqn}), which gives
\begin{equation}
\begin{split}
    \tau_{\ell n} = \frac{r^3}{2} \Bigg\{ \left[f_\ell'(k_{\ell n}r)\right]^2 &- \frac{f_\ell(k_{\ell n}r)f_\ell'(k_{\ell n}r)}{k_{\ell n}r} \\ 
    &- f_\ell(k_{\ell n}r)f_\ell''(k_{\ell n}r) \Bigg\}\Bigg|_{r_1}^{r_2}\,.
\end{split}
\end{equation}
Now we could explicitly write down the following normalization factors for different BCs.
\begin{itemize}
    \item For a \textit{sphere}, $r_1=0$, $f_\ell=j_\ell$,
    \begin{itemize}
        \item with \textit{Dirichlet} BC,
        \begin{equation}
            \tau_{\ell n} = \frac{r_2^3}{2} [j_{\ell+1}(k_{\ell n}r_2)]^2 \,.
        \end{equation}
        \item with \textit{Neumann} BC,
        \begin{equation}
            \tau_{\ell n} = \frac{r_2^3}{2} \left[ 1-\frac{\ell(\ell+1)}{(k_{\ell n}r_2)^2} \right] [j_\ell(k_{\ell n}r_2)]^2 \,.
        \end{equation}
    \end{itemize}
    \item For a \textit{shell}, $0<r_1<r_2$ and $f_\ell=\mathcal{J}_\ell$,
    \begin{itemize}
        \item with \textit{Dirichlet} BC,
        \begin{equation}
            \tau_{\ell n} = \left. \frac{r^3}{2} [\mathcal{J}_{\ell+1}(k_{\ell n}r)]^2\ \right|_{r_1}^{r_2} \,.
        \end{equation}
        \item with \textit{Neumann} BC,
        \begin{equation}
            \tau_{\ell n} = \left. \frac{r^3}{2} \left[ 1-\frac{\ell(\ell+1)}{(k_{\ell n}r)^2} \right] [\mathcal{J}_\ell(k_{\ell n}r)]^2\ \right|_{r_1}^{r_2} \,.
        \end{equation}
    \end{itemize}
\end{itemize}

\section{\label{sec:sfb_3d}SFB and 3D Cartesian power spectra}
Here we discuss the relation between SFB and 3D Cartesian power spectra.
Assuming that $f(\bm{r})$ is a statistically homogeneous and isotropic 3D field, whose auto-power spectrum in Cartesian coordinates is given through
\begin{equation}
    \langle f(\bm{k})f^*(\bm{k}') \rangle = (2\pi)^3 \deltaD(\bm{k}-\bm{k}') P_{f}(k) \,.
\end{equation}
The SFB coefficient with or without boundary conditions (BCs) can be written in a general form as
\begin{equation} \label{eq:sfb_general_f}
    f_{\ell m}(k_r) = \int dr\, r^2 \mathcal{F}(k_r,r) \int d\Omega\, f(\bm{r}) Y^*_{\ell m}(\hat{r}) \,,
\end{equation}
where $\mathcal{F}(k_r,r)$ includes the factor and radial eigenfunction, and $k_r$ denotes the radial wavenumber in SFB basis.
By transforming $f(\bm{r})$ to $f(\bm{k})$ and using the plane wave expansion in Eq.~(\ref{eq:plane_wave_expansion}), the inner angular integral can be written as an integral over the 3D wavevector
\begin{equation} \label{eq:angular_3Dk}
    \int d\Omega\, f(\bm{r}) Y^*_{\ell m}(\hat{r}) = \frac{i^\ell}{2\pi^2}\int d^3k \, f(\bm{k}) j_\ell(kr) Y^*_{\ell m}(\hat{k}) \,,
\end{equation}
and Eq.~(\ref{eq:sfb_general_f}) becomes
\begin{equation} \label{eq:sfb_int_k}
    f_{\ell m}(k_r) = \frac{i^\ell}{2\pi^2}\int d^3k \, f(\bm{k}) Y^*_{\ell m}(\hat{k}) \int dr\, r^2 \mathcal{F}(k_r,r) j_\ell(kr) \,.
\end{equation}

Without any BC, $k_r$ is continuous and
\begin{equation}
    \mathcal{F}(k_r,r) = \sqrt{\frac{2}{\pi}}k j_\ell(k_r r) \,,
\end{equation}
where the choice of the normalization factor is not important here and may vary depending on the convention.
Then the integral over $r$ from 0 to $+\infty$ gives $\deltaD(k_r-k)$ and the auto-correlation turns out to be
\begin{equation}
    \langle f_{\ell m}(k_r) f^*_{\ell' m'}(k_r') \rangle = \deltaK_{\ell\ell'}\deltaK_{m m'}\deltaD(k_r-k_r')P_f(k_r) \,,
\end{equation}
i.e. we have $C^f_\ell(k_r)=P_f(k_r)$ and the radial wavenumber is exactly the 3D wavenumber.

With a shell or sphere BC, $k_r$ are discrete $k_{\ell n}$ values and we have
\begin{equation}
    \mathcal{F}(k_{\ell n},r) = \tau^{-1}_{\ell n} \mathcal{J}_\ell(k_{\ell n}r) \,.
\end{equation}
The auto-correlation now reads
\begin{equation}
\begin{split}
    \deltaK_{\ell \ell'}\deltaK_{m m'} C^f_{\ell n n'} =&\ \langle f_{\ell m}(k_{\ell n}) f^*_{\ell' m'}(k_{\ell' n'}) \rangle \\
    =&\ \deltaK_{\ell \ell'}\deltaK_{m m'} \frac{2}{\pi} \tau^{-1}_{\ell n}\tau^{-1}_{\ell n'} \\
    & \times \int dk\, k^2 P_f(k) I_{\ell n}(k) I_{\ell n'}(k) \,,
\end{split}
\end{equation}
where we defined
\begin{equation}
    I_{\ell n}(k) \equiv \int_{r_1}^{r_2} dr\, r^2 \mathcal{J}_\ell(k_{\ell n}r) j_\ell(kr) \,.
\end{equation}
Since $k$ is arbitrary, this integral over $r$ would no longer reduces to the Delta function.

\section{\label{sec:Gaussian_cov_ps}Gaussian covariances between power spectra}
In this Appendix, we briefly discuss the Gaussian sampling covariances for SH and SFB power spectra based on the pseudo-$C_\ell$ (PCL) estimator.
Note that although the expressions below are written for 3D SFB coefficients, the derivation is the same for SH coefficients and thus any 3D field $f(r,\hat{r})$ can be replaced with 2D field $a(\hat{r})$ by simply erasing the corresponding radial wavenumber index $n$.

The PCL estimator is constructed based on the equivalence of all the $m$ modes
\begin{equation} \label{eq:pseudo-cl}
    \hat{C}_{\ell n n'}^{ff'} = \frac{1}{(2\ell+1)f_{\rm sky}} \sum_{m=-\ell}^\ell \hat{f}^*_{\ell m n} \hat{f}'_{\ell m n'} \,,
\end{equation}
i.e. the estimate for each $\ell$ mode is given by the average over all the $2\ell+1$ $m$ modes,  $f_{\rm sky}$ is the fractional sky coverage and we ignore the coupling between multipoles for simplicity, as also assumed in the main text.
Assuming the fields $f_i$ to be Gaussian, with Wick contraction
\begin{equation}
\begin{split}
    \langle f_1 f_2 f_3 f_4 \rangle = &\langle f_1 f_2 \rangle \langle f_3 f_4 \rangle \\
    &\ + \langle f_1 f_3 \rangle \langle f_2 f_4 \rangle + \langle f_1 f_4 \rangle \langle f_2 f_3 \rangle \,,
\end{split}
\end{equation}
we can show that the sample covariance reads
\begin{equation} \label{eq:Gaussian_cov_ps}
\begin{split}
    &{\rm Cov}\left( \hat{C}_{\ell n_1 n_2}^{f_1 f_2}, \hat{C}_{\ell' n_3 n_4}^{f_3 f_4} \right) \\
    &\,= \left\langle \left(\hat{C}_{\ell n_1 n_2}^{f_1 f_2} - C_{\ell n_1 n_2}^{f_1 f_2}\right) \left(\hat{C}_{\ell' n_3 n_4}^{f_3 f_4} - C_{\ell' n_3 n_4}^{f_3 f_4}\right) \right\rangle \\
    &\,= \frac{\deltaK_{\ell\ell'}}{(2\ell+1)f_{\rm sky}} \left( C_{\ell n_1 n_3}^{f_1 f_3} C_{\ell n_2 n_4}^{f_2 f_4} + C_{\ell n_1 n_4}^{f_1 f_4} C_{\ell n_2 n_3}^{f_2 f_3} \right) \,,
\end{split}
\end{equation}
where $C_{\ell n_1 n_2}^{f_1 f_2}$ are measured power spectra that include possible noises.
As mentioned in the main text, in this work we consider the lensing reconstruction noise and shot noise in CMB lensing and galaxy clustering auto-power spectra respectively.
Besides the simplified $f_{\rm sky}$ description of partial sky coverage, \cite{Tristram2005} presented a full discussion of the PCL estimator and also the corresponding Gaussian covariance matrix for SH power spectra with angular masks included.
Similar discussion and expressions should also work for the joint analyses of SH and SFB since there is no difference in their angular multipole descriptions, see e.g.~\cite{Pratten2013,Lanusse2015} for discussions about the impact on SFB power spectra.
A recent work on SFB power spectrum estimator~\cite{Gebhardt2021} also presents more detailed discussions about the analytical covariance matrix.

\section{\label{sec:num_cal}Numerical computation of power spectra}
This Appendix includes some details on the numerical computation of the power spectra.
In our formalism discussed in Section~\ref{subsec:power_spectra}, there are mainly two steps.
First, for tracers of the matter field, e.g. CMB lensing and 2D projected or 3D galaxy overdensity, we need to the compute their transfer functions given by the line-of-sight integrals over radial distance $r$.
Then these transfer functions can be combined with the matter power spectrum in the integral over 3D wavenumber $k$ to get the SH or SFB power spectra.
The numerical evaluation of the line-of-sight integrals is nontrivial given the highly oscillatory $j_\ell(kr)$ functions, and the Limber approximation~\cite{Limber1953,LoVerde2008}
\begin{equation} \label{eq:limber}
\begin{split}
    \int dx\,f(x)j_\ell(x) \simeq \int dx\,f(x)&\sqrt{\frac{\pi}{2\ell+1}} \\
    &\times\deltaD(\ell+1/2-x)
\end{split}
\end{equation}
is usually used to speed up the computation.
However, this approximation work well only for high $\ell$s, an integral range that is much wider than the oscillation period of $j_\ell(x)$, and also $f(x)$ should vary slowly compared with $j_\ell(x)$.
These requirements may not hold in our case.
First, we are interested in large scales and very low $\ell$s are included.
Also, the redshift slice (i.e. the integral range) could be very narrow given the large number of bins in TSH analysis.
Besides, for SFB analysis, we have the radial basis function $\mathcal{J}_\ell(k_{\ell n}r)=j_\ell(k_{\ell n}r)+A_{\ell n}y_\ell(k_{\ell n}r)$ in the line-of-sight integral, which is also oscillating very fast as $j_\ell(kr)$.
Given these issues, we are not able to use Limber approximation in our analyses.
Instead, we evaluate the integral in a brute-force but exact way with a large number of sampling points.
A significant fraction of time is spent on getting the spherical Bessel functions $j_\ell(kr)$ or $\mathcal{J}_\ell(k_{\ell n}r)$, which are computed recursively.
To speed this up, we tabulate these on the 2D $(k,r)$ or $(k_{\ell n},r)$ sampling grids for each $\ell$ in advance, which can then be loaded wherever needed.
In Fisher analysis, since we only need to evaluate these power spectra a few times, the computational time is acceptable.
But for a MCMC fitting of the analytic power spectra to the estimates from real data, it would be necessary to make some improvements since the brute-force computation is too slow for each step in the MCMC chains.
For angular power spectra, the \textsc{FFTLog} algorithm~\cite{Hamilton2000} has been used to optimize the computation, see e.g.~\cite{Assassi2017,Chen2021}.
For future work, it would be useful to check if this algorithm could also be applied to the SFB power spectra given the radial basis functions in the integral.

\bibliography{lensingxgal_sfb}

\end{document}